\documentclass[10pt,letterpaper]{article}
\usepackage[top=0.85in,left=2.75in,footskip=0.75in]{geometry}

\usepackage{amsmath,amssymb}
\usepackage{changepage}
\usepackage{textcomp,marvosym}
\usepackage{cite}
\usepackage{nameref,hyperref}

\usepackage{xspace}
\usepackage{lipsum}
\usepackage{float}
\usepackage{caption}
\usepackage{subcaption}
\usepackage{tabularx}
\usepackage{longtable}
\usepackage{booktabs}
\usepackage{amssymb}
\usepackage{amsmath}
\usepackage{hyperref}
\usepackage{changepage}
\usepackage[numbers,square]{natbib}
\usepackage[T1]{fontenc}
\usepackage{lmodern}

\usepackage[right]{lineno}
\usepackage[nopatch=eqnum]{microtype}
\DisableLigatures[f]{encoding = *, family = * }
\usepackage[table]{xcolor}
\usepackage{array}
\newcolumntype{+}{!{\vrule width 2pt}}
\newlength\savedwidth

\raggedright
\usepackage[aboveskip=1pt,labelfont=bf,labelsep=period,justification=raggedright,singlelinecheck=off]{caption}

\makeatletter
\renewcommand{\@biblabel}[1]{\quad#1.}
\makeatother
\usepackage{lastpage,fancyhdr,graphicx}
\usepackage{epstopdf}
\fancyheadoffset[L]{2.25in}
\fancyfootoffset[L]{2.25in}
\begin{document}
\vspace*{0.2in}

\begin{flushleft}
{\Large
\textbf\newline{Remote sensing for sustainable river management: Estimating riverscape vulnerability for Ganga, the world’s most densely populated river basin} 
}
\newline
\\
Anthony Acciavatti\textsuperscript{1,2*},
Sarthak Arora\textsuperscript{2},
Michael Warner\textsuperscript{2},
Ariel Chamberlain\textsuperscript{2},
James C. Smoot\textsuperscript{2},
Nikhil Raj Deep\textsuperscript{2,3}, 
Claire Gorman Hanly\textsuperscript{2,4}
\\
\bigskip
\textbf{1} Yale School of Architecture, 180 York St, New Haven, CT, 06511, United States
\\
\textbf{2} Collaborative Earth, 1080 Jones St, Berkeley, CA, 94710, United States
\\
\textbf{3} University of Florida IFAS Department of Soil, Water, and Ecosystem Sciences, 2181 McCarty Hall A, Gainesville, FL, 32611, United States
\\
\textbf{4} MIT School of Architecture and Planning, 77 Massachusetts Ave building 10-400, Cambridge, MA, 02139, United States
\\
\bigskip

* anthony.acciavatti@yale.edu

\end{flushleft}
\section*{Abstract}
Surface water mixed with wastewater creates serious environmental concerns, particularly in densely populated urban areas with inadequate infrastructure. Such contamination threatens to cause major public health crises in the Ganga Basin where monsoonal flooding converges with 6 billion liters of untreated sewage that is discharged daily into the basin by 650 million people. GIS-based analytic hierarchy process (AHP) with remote sensing data was conducted to highlight areas of vulnerability along a 20-km wide riverscape. Analytic network process (ANP), Nested AHP, fuzzy AHP, and 1-N AHP (novel variant of AHP) were used to constrain AHP model uncertainties, and composites of these analyses were utilized to define the vulnerability of the river Ganga to pollution. AHP categorized 83.7\% of the area as having extremely low or low vulnerability and 3.5\% of the area as having highly or extremely high vulnerability. ANP and Nested AHP produced focused, yet dampened, vulnerability-score maps compared to AHP. Fuzzy AHP and 1-N AHP detected sensitivities to factor variability and potential unknown acute and chronic factors. While fuzzy AHP identified quintile-level changes in vulnerability based on scenario parameters, vulnerability scores of 1-N AHP and AHP showed no major differences. Normalized composite vulnerability \(\geq\) 2 standard deviations highlighted particularly vulnerable locations and identified instances where network effects were greater than factor class and vice versa. Together, these analyses located areas of extreme vulnerability at the nexus of river Ganga and urban landscapes as well as regions of low vulnerability potentially suitable for conservation efforts or sustainable development practices to prevent their degradation. This approach contributes to a more comprehensive understanding of remote sensing data applications in environmental assessment, and these decision-making variants can also have broader applications in other areas of environmental management and sustainability, facilitating more precise and adaptable decision support frameworks in densely populated watersheds.


\section*{Introduction}

India is home to the world’s largest population and most populated river basin: Ganga. While it spans across Bangladesh, Nepal, and Tibet, the majority of the Ganga Basin and its population resides within present-day India. Not only is the Ganga Basin densely populated, but it also remains agriculturally productive and receives significant rainfall during the southwest monsoon season (June-September); all three factors compound its vulnerability to pollution and disease vectors. 

The acute and chronic impacts of human activities on the river Ganga are physical, chemical, and biological~\citep{24,12}. For instance, the Ganga Basin is physically transformed by several barrages (i.e., dams) as well as an extensive irrigation system that diverts the Himalayan discharge away from river Ganga to support agriculture through the basin. Moreover, changes in water chemistry along the river are driven by a variety of factors, including industrial waste and synthetic fertilizer runoff, along the length of the river~\citep{23}. Additionally, Srinivas et al. \cite{26} identified 13 threats and challenges impacting the river Ganga that included broad concerns ranging from biodiversity loss to open defecation near the river. To reduce anthropogenic induced threats and increase the overall health of the river basin, river rejuvenation and development must simultaneously address multiple social, economic, and environmental dimensions~\citep{16}.

In order to prioritize areas of the Ganga Basin for actions such as conservation, restoration, intervention, and rejuvenation, it is crucial to perform vulnerability assessments. There are many vulnerability assessment methods, and a multitude of factors, dimensions, and subdimensions may be included in an assessment~\citep{17}. The Analytical Hierarchical Process (AHP) was used because our aim was to produce an approach that properly recognizes that the interaction of pollutants with local geophysical factors creates variation in environmental impacts downstream. Further, our intent was to identify not only the immediate vulnerability of a given area but also the relative threat to the broader river system posed by potential sources of pollution in that area. This was done rather than generating a water management or environmental impact assessment decision such as realized by PROMETHEE or TOPSIS.

AHP is a multicriteria decision making process that combines  pairwise comparisons and expert judgements to arrive at its conclusions. Developed by Saaty \cite{20}, AHP has been used in numerous applications, ranging from resource allocation and site prioritization to risk assessment and conflict resolution. Climate and geophysical models have proven well-suited for AHP, due to the complexity of decision making under uncertainty with respect to factors impacting their features of interest. For example, Jhariya et al. \cite{8} used AHP in combination with Geospatial Information Systems (GIS), remote sensing, and vertical electrical soundings (VES) to identify potential zones of groundwater using weighted data on geology, geomorphology, rainfall, lineament, LULC (land use and land cover), drainage density, slope, soil type, and soil texture. Using AHP, they were able to estimate the likelihood of groundwater in each zone, with five levels of potential (low, medium, medium-high, high, and very high) with an accuracy of 80\%.

\subsection*{AHP factors of waterway vulnerability and their Saaty rankings}

In this study, AHP was used as a characterization scheme rather than a decision making process; hence, factors and their relative rank order of importance were structured to highlight areas that may be vulnerable to pollution with an emphasis on urban pollution. Land use was identified using LULC, a widely recognized direct characterization of land types and includes development such as urbanization and agricultural uses. Urbanization may be directly related to major sources of contamination in water bodies, while agriculture is a major source of nonpoint source pollution including fertilizers and pesticides. Moreover, densely populated areas have a direct impact on surface water quality. For instance, population growth rate and water quality parameters, such as biochemical oxygen demand (BOD) and dissolved oxygen, were highly correlated in a Kelani river watershed in Sri Lanka~\citep{9}. Given our emphasis on urban pollution and its direct impact on vulnerability, population density (PD) was ranked as the most important factor (Saaty-scale value of 1), followed by LULC (Saaty-scale value of 2).

Rainfall, slope, and drainage density indirectly impact vulnerability in that they are not sources of pollution, yet they directly influence erodibility and pollution mobility. For example, the runoff from steeper slopes was more likely to carry pollutants into streams than runoff from land use in flatter slopes~\citep{30}, and effects on water quality originating from land use (agriculture, industrial, and residential) adjacent to water bodies were dependent on rainfall variability~\citep{15}. Drainage density (DD) is an indicator of surface runoff processes and direct runoff and pollution transport increase with greater DD~\citep{3}. Rainfall, DD, and slope were assigned Saaty-scale values of 4, 5, and 7 respectively because rainfall initiates pollutant movement events, slope is an important factor in pollutant transport, and DD influences final distribution of pollutants.

Temperature is an environmental determinant of microbial activity and productivity including nuisance algal blooms. For instance, water surface temperature behavior in zones with algal bloom occurrences presented greater significant values, up to 3°C, than those with clearer water~\citep{6}. Moreover, seasonal increases of anaerobic bacteria in aquatic surface sediment during summer compared to cool months indicated a depletion of oxygen in the overlying water~\citep{25}. Given the absence of high resolution site monitoring throughout the Ganga Basin, land surface temperature (LST) was used as a surrogate temperature estimate. LST was assigned the lowest importance among all input factors (9 on the Saaty-scale) because it was expected to have a minor overall relative impact on vulnerability.

\subsection*{Potential Shortcomings of AHP}

While AHP offers numerous advantages in terms of decision making in remote sensing, Munier and Hontoria \cite{13} identified 30 potential shortcomings of the method. Given the importance of accurately understanding vulnerability, we reviewed all the shortcomings of AHP and assessed whether they were applicable for remote sensing data and our particular use case (\nameref{S1_Table}). For instance, the first critique “The Pair-Wise Method and Its Application in AHP” suggested that some problems may not be suited for AHP if their decision space is amorphous with many interrelated factors that have minimal differences in their relative importance. This critique was not applicable for our use-case, since our problem fits well to the AHP architecture. Similarly, the critique around rank reversals, which arise when uncertainty in the decision space is larger than the consistency of rank, was applicable on our use-case since a change in parameter weights might lead to an inaccurate assessment of the vulnerability of the river. To test for this and design a better method, we applied AHP while taking into account the effect of unknown variables (1-N AHP) and Fuzzy AHP as diagnostic analyses. To summarize, all the applicable shortcomings were addressed with the following approaches: Analytic Network Process (ANP)~\citep{22}, Fuzzy AHP~\citep{21} , AHP with non-linear parametric influence (Nested AHP), 1-N AHP, and Fuzzy AHP while taking into account the effect of unknown factors (Fuzzy 1-N AHP) (\nameref{S1_Table}).

In this paper, we apply  AHP to our remote sensing datasets to assess the vulnerability of the river Ganga to pollution along a 1,330 km stretch of the river. We cover the problem formulation, results and inferences of AHP. We also include an analysis of each shortcoming of AHP in the context of remote sensing, addressing the list set forth by Munier and Hontoria \cite{13}. We share whether each shortcoming was applicable to our problem statement, while suggesting and attempting alternative solutions to counteract the shortcomings. Through this dialogical process, we have arrived at  two novel decision making approaches inspired by AHP, namely 1-N AHP and Fuzzy 1-N AHP. The methodological contributions of this research are, at one level, the utilization of AHP variants to constrain process uncertainties; and at another level, the demonstration of how incremental evaluation of the AHP process and its shortcomings relative to a given problem can reveal application-specific permutations of the original (i.e., a variant (or a combination of variants) might work better for one problem, while another variant(s) might work for another problem, depending upon the most important shortcomings a decision maker might want to address in their use-case). Finally, we report a pollution vulnerability evaluation of the river Ganga along a 20-km wide riverscape produced by these processes (e.g., AHP provided an overall assessment of vulnerability, ANP and Nested AHP further highlighted areas of potential interest, and fuzzy analyses showed factor sensitivity to rank reversal as well as locations sensitive to shifts in vulnerability due to factor weight variability). 

\section*{Methods and Data}

\subsection*{Study area and data source}

\begin{figure}[h! tbp]
\begin{center}
    \includegraphics[width=0.7\textwidth]{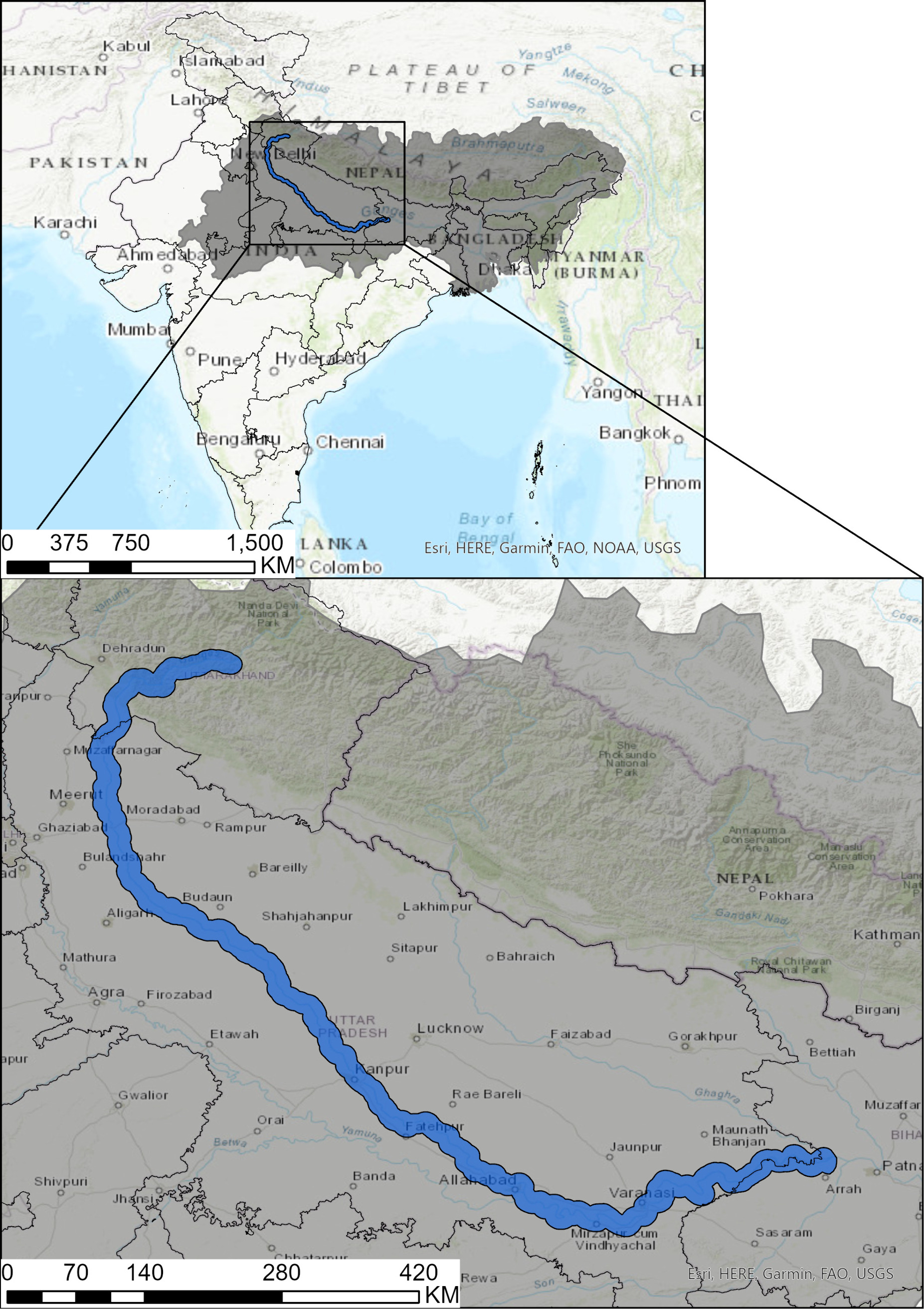}    
\end{center}
\caption{{\bf Ganga Inset Map}
This map shows the Ganga Basin, in gray, and the buffer region, in blue,  which is the scope of our paper. The 20 km buffer zone extends 10 km on both sides of the main channel of river Ganga.}
\label{fig1}
\end{figure}

The area of analysis (26,609.4 km\textsuperscript{2}) is a 20 km zone extending 10 km on both sides of the main channel of river Ganga from Haridwar, Uttarakhand, where it makes its debouche from the Siwalik Hills into the Indo-Gangetic plains to its confluence with river Ghaghara at Sitab Diara, Uttar Pradesh, India (Fig~\ref{fig1}). 

\begin{figure}[h! tbp]
\begin{center}
    \includegraphics[width=0.7\textwidth]{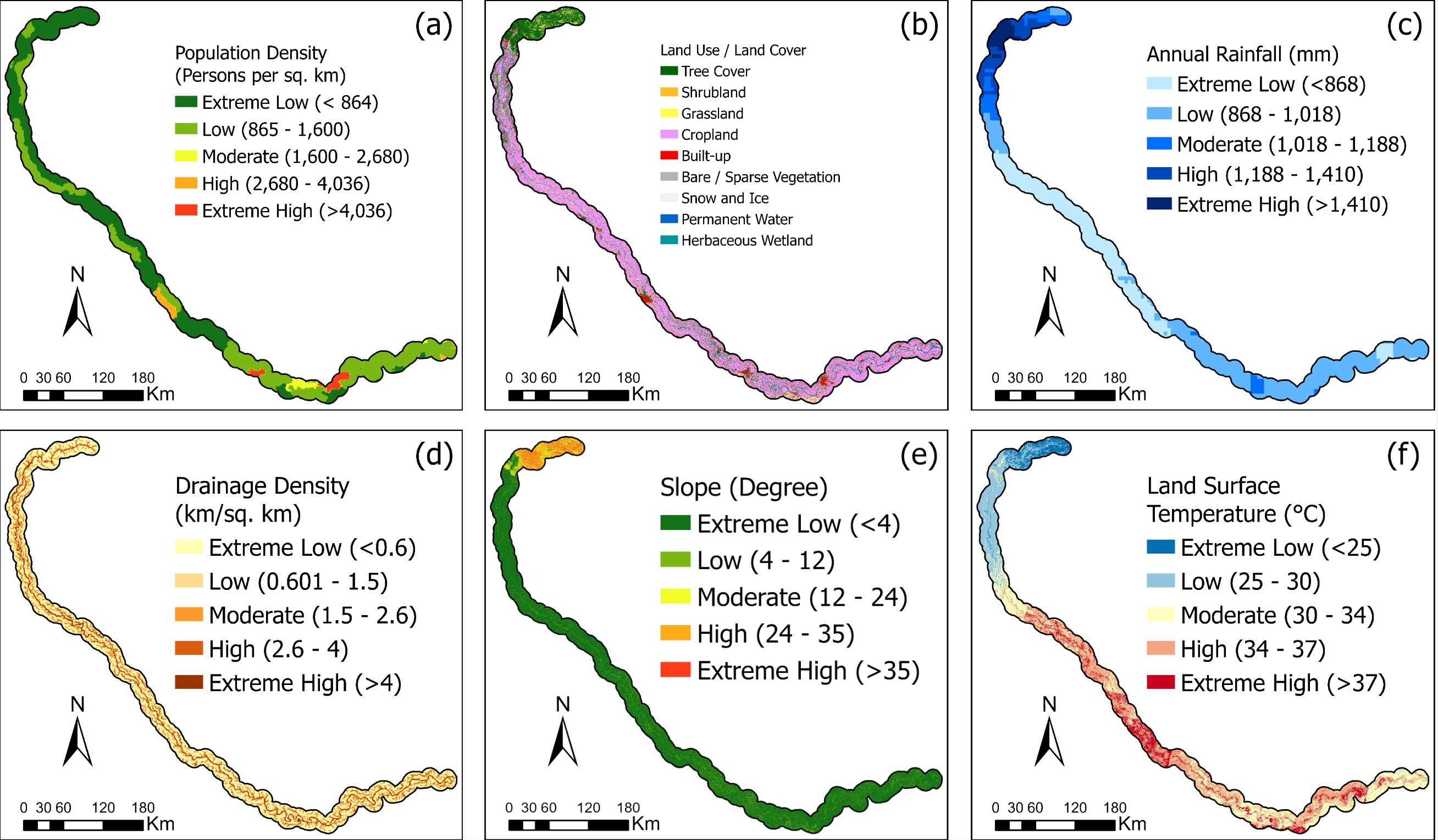}    
\end{center}
\caption{{\bf AHP Factor Layers}
Visual representation of the data used in the calculation of AHP. AHP factors Population Density (a), Annual Rainfall (c), Drainage Density (d), Slope (e), and Land Surface Temperature (LST) (f) were divided into extreme low to extreme high using natural breaks. AHP factor Land Use Land Cover (LULC) (b) was separated by land use class. Factor classes were based on natural breaks. Factor classification indicates how much impact each variable has on vulnerability.}
\label{fig2}
\end{figure}

The change in elevation along this 1,330 km stretch of Ganga is from 314 m to 76 m above mean sea level. Data from Landsat-8 (Level 2, Collection 2, Tier 1; pixel resolution, 30x30m) and other sources (\nameref{S2_Table}) were used to assess vulnerability in the area of analysis include slope, drainage density (DD), land surface temperature (LST), rainfall, land use/land cover (LULC), and population density (Fig~\ref{fig2}). The drainage density and center line for the river Ganga were both derived from a digital elevation model (DEM) of the watershed using typical GIS hydrography processes, as previously described by Rahaman et al. \cite{19}. Streams were extracted from NASADEM data as a vector format Shapefile using the R programming language, Whitebox, Terra, and sf packages. Lower order streams were removed to obtain the main Ganga river centerline with ArcGIS Pro. The Ganga center line down-river of the Uttar Pradesh border was removed, and the 20 km zone was established as a buffer around this shortened centerline within the river Ganga. 

\subsection*{Analytical Hierarchy Process (AHP)}

To categorize the vulnerability of the area immediately around the river Ganga and to identify potential locations to implement our remediation strategies, we applied the AHP on remote sensing and other geospatial data as previously described by Jhariya et al. \cite{8}. Environmental, topographical, and anthropogenic factors were ranked from equal importance to relatively greatest hierarchical importance across the full breadth of the 9-point Saaty scale based on their likelihood of impacting vulnerability. A pairwise matrix constructed from these ranked factors and their reciprocal values were used to estimate a mean normalized weight for each factor (\nameref{S3_Table}). These normalized weights were applied to each of five factor classes that were identified by natural breaks in the data using the Jenks method. LULC categories were numerically scored as vulnerability criteria based on their linkage to pollution. Lastly, overall vulnerability per pixel was determined from the sum of the factor vulnerability scores. Additional details describing factor rank orders, pairwise comparison matrices and their weights, as well as Jenks natural breaks are provided in Supporting Information (\nameref{S1_Text} and \nameref{S2_Text}).


If:
\begin{itemize}
\item \(W\) is the 1x6 row vector of AHP weights:
\begin{equation}\label{eq1}
W = \begin{bmatrix} w_1 & w_2 & w_3 & w_4 & w_5 & w_6 \end{bmatrix}
\end{equation}
					
\item \(X(p)\) is the 1x6 row vector of parameter values at pixel p for the six factors:
\begin{equation}\label{eq2}
X(p) =\begin{bmatrix} x_1(p) & x_2(p) & x_3(p) & x_4(p) & x_5(p) & x_6(p) \end{bmatrix}
\end{equation}

\end{itemize}
			
We calculated the predicted vulnerability for each pixel by multiplying each factor's parameter value by its corresponding AHP weight i.e., the dot product of the vector \(W\) with the transpose of the vector \(X(p)\):
\begin{equation}\label{eq3}
V(p) = W \cdot X(p)^T										
\end{equation}

Eq~(\ref{eq1}),~(\ref{eq2}) and (\ref{eq3}) were used to create the vulnerability map of the Ganga buffer region. We used Google Earth Engine to multiply the AHP weights with the bucketed values for each layer at each pixel and aggregated them to get our final vulnerability scores for the entire Ganga buffer. More information about the remote sensing datasets we used on Google Earth Engine can be found in \nameref{S2_Table}. 

\subsection*{Diagnostic Analyses}

\subsubsection*{Nested AHP}

We performed the AHP analysis under the assumption that while each parameter contributes to the vulnerability differently, the contribution of each parameter to vulnerability was linear in nature i.e., areas where the value of a parameter was 5 contributed to the vulnerability 5 times more than areas where the value was 1. This is an overly simplified approach; hence, we created a pairwise matrix for each factor, using the bucketed values for the parameters, and performed AHP on each parameter with itself, to create the Nested AHP approach (\nameref{S3_Text}).

\subsubsection*{Fuzzy AHP}

Fuzzy Analytic Hierarchy Process (Fuzzy AHP), as introduced by Saaty \cite{21} is an extension of the classic AHP that performs a sensitivity analysis on the AHP results by incorporating the fuzziness (probabilistic alterations/error) associated with human judgment. By using fuzzy numbers in the rankings, Fuzzy AHP accommodates these uncertainties and thus provides more reliable results. To understand the robustness of our AHP factor rank order results, we utilized fuzzy AHP to see if there was an inconsistency within the results, i.e., if there was a rank reversal between the parameters when we incorporate fuzziness to the values of the AHP pairwise comparison matrix (\nameref{S4_Text}). 

\subsubsection*{ANP}

One of the main limitations of AHP is its inability to adequately capture the interdependencies among criteria and alternatives. To address this issue, Saaty \cite{22} developed ANP, which extends the AHP methodology by accommodating feedback loops and interdependencies among criteria and alternatives. Unlike AHP, which assumes a hierarchical structure, ANP allows for the representation of complex relationships and feedback loops in decision networks. The method utilizes a supermatrix to integrate both local and global influences, providing a more realistic representation of environments. ANP is widely used to represent interdependent relationships, making it suitable for decision problems with interconnected elements. ANP introduces the concept of contextual weights, enabling decision makers to adjust the importance of criteria and alternatives based on the context of the decision problem. This flexibility enhances the adaptability of ANP to various situations. While there is undoubtedly overlap with Nested AHP, ANP attempts to include marginal effects between factors whereas Nested AHP compounds the effects of classes.

In a complex system such as the geophysical environment, it is expected that AHP factors will be correlated to one another. While AHP assumes strict independence between criteria at different levels of the hierarchy, ANP allows for both dependence and independence relationships (Table~\ref{table2}) as described previously by Poh and Liang \cite{18}; this flexibility better captures the intricacies of decision networks across natural systems.

\begin{table}[H]
\centering
\renewcommand{\arraystretch}{1.2} 
\caption{\bf Vulnerability ANP Supermatrix\textsuperscript{a}}
\begin{tabular}{@{}|p{5cm}|p{1.25cm}|p{1.25cm}|p{1.75cm}|p{2cm}|@{}}
\hline
\textbf{} & 	\textbf{Goal} & 	\textbf{Criteria} & 	\textbf{Subcriteria} & 	\textbf{Alternative outcomes} \\ 
\hline
Goal & 	0 & 	0 & 	0 & 	0 \\ \hline
Criteria (AHP factors) & 	W21 & 	W22 & 	0 & 	0 \\ \hline
Subcriteria (vulnerability classes) & 	0 & 	W32 & 	W33 & 	0 \\ \hline
Alternative outcomes & 	0 & 	0 & 	W34 & 	I \\ \hline
\end{tabular}
\vspace{0.5em}
\\
\begin{minipage}{0.75\textwidth}
    \footnotesize\textsuperscript{a} ANP Criteria were AHP factors, and ANP Subcriteria were AHP vulnerability rankings. \\
    The ANP supermatrix component matrices as follows:
    \begin{itemize}
        \item Relative importance of criteria (AHP factors), \(W_{21}\);
        \item Inner dependence (As defined by \cite{18}) matrix of criteria, \(W_{22}\);
        \item Relative importance of subcriteria (vulnerability classes) w.r.t criteria, \(W_{32}\);
        \item Inner dependence matrix of subcriteria, \(W_{33}\);
        \item and Relative importance of subcriteria w.r.t alternatives, \(W_{34}\). \\
    \end{itemize}
    \[
    \begin{aligned}
        W_{21} &= \begin{bmatrix}
        0.408 \\
        0.268 \\
        0.138 \\
        0.1 \\
        0.058 \\
        0.028 \\
        \end{bmatrix} \quad &
        W_{22} &= \begin{bmatrix}
        0.464 & 0.233 & 0.14 & 0.277 & 0 & 0.186 \\
        0.271 & 0.408 & 0.12 & 0.098 & 0 & 0.077 \\
        0.105 & 0.126 & 0.662 & 0 & 0 & 0 \\
        0.085 & 0.079 & 0 & 0.498 & 0 & 0 \\
        0.076 & 0.079 & 0.078 & 0.126 & 0 & 0 \\
        0 & 0.075 & 0 & 0 & 0 & 0.737 \\
        \end{bmatrix} 
    \end{aligned}
    \]
    \[
    \begin{aligned}
    W_{32} &= \begin{bmatrix}
        0.572 & 0.505 & 0.501 & 0.642 & 0.388 & 0.557 \\
        0.207 & 0.186 & 0.172 & 0.157 & 0.233 & 0.165 \\
        0.114 & 0.144 & 0.124 & 0.071 & 0.276 & 0.129 \\
        0.057 & 0.11 & 0.102 & 0.065 & 0.07 & 0.099 \\
        0.05 & 0.055 & 0.102 & 0.065 & 0.033 & 0.05 \\
        \end{bmatrix} \quad & 
        W_{33} = O_{5x5}
    \end{aligned}
    \]
    \[
    \begin{aligned}
        W_{34} &= \begin{bmatrix}
        0.56 & 0.61 & 0.571 & 0.483 & 0.409 \\ 
        0.187 & 0.203 & 0.229 & 0.276 & 0.273 \\ 
        0.112 & 0.102 & 0.114 & 0.138 & 0.182 \\ 
        0.08 & 0.051 & 0.057 & 0.069 & 0.091 \\ 
        0.062 & 0.034 & 0.029 & 0.034 & 0.045 \\ 
        \end{bmatrix} \qquad & 
        I = I_{5x5} \\
    \end{aligned}
    \]
        
    The calculation of each component matrix is available in the “ANP supermatrix and inner dependence pairwise comparisons” section in the Supplemental Information (\nameref{S5_Text}).
    \end{minipage}
\label{table2}
\end{table}

We generated the ANP supermatrix with the goal to assign pollution vulnerability scores (PVS). Here the AHP Factors, or Criteria were - Population density, LULC, rainfall, drainage density, slope and temperature. The Alternatives were pertaining to the PVS, i.e., extremely high, high, moderate, low, extremely low. The possible relationship arcs were dependencies, self-loops, and feedbacks. We used the Python library pyanp created by Adams et al \cite{1} to run the ANP Algorithm on the Supermatrix.

\subsubsection*{1-N AHP}

Another shortcoming of AHP is that, from inside the process, there is no way of knowing for certain if there are any important factors missing from the decision problem parameters. For AHP, our model included six factors: population density, LULC, rainfall, drainage density, slope and temperature. If a fundamental pollution vulnerability factor was missing, AHP would not capture it. To overcome this, we devised an approach named 1-N AHP where we aimed to validate whether the unknown and unconsidered factors are important for our analysis of vulnerability or not (\nameref{S6_Text}).

\subsubsection*{Fuzzy 1-N AHP}

We have described two approaches to deal with the complex nature of decision making methods for environmental problems. While Fuzzy AHP took into account the potential error in the decision maker's choice for the values of the pairwise comparison matrix, 1-N AHP considered unknown factors which may have been missed while considering the vulnerability of a river. Since both approaches were mutually exclusive, we performed one final test to check the robustness of our results by combining these two approaches.

We performed 1-N AHP followed by a fuzzy analysis on the values. First, we used our eight weights (six original weights, plus a weight each for aggregated acute and chronic factors) that we derived from our analysis in 1-N AHP, and applied linear algebra to create our new matrix (i.e., because the AHP vector output is the eigenvector, and the eigenvalue is used to calculate the confidence, we used \(A = PDP^{-1}\) to get back our original matrix \(A\), where \(P\) is a matrix of eigenvectors and \(D\) is a diagonal matrix of eigenvalues). We then ran Fuzzy AHP on this new 8x8x3 matrix for 100,000 simulations to get the 8x1 size priority vectors. From these vectors, we randomly excluded the acute factors for 97.5\% of the cases, and in 75\% of the 100,000 cases, the chronic factors were randomly excluded, before checking the rank reversals. This was to account for the probabilities of the presence (or absence) of acute and chronic factors respectively while checking for rank reversals. 

\section*{Results}

\subsection*{River Ganga vulnerability based on AHP}

\begin{figure}[h! tbp]
\begin{center}
    \includegraphics[width=0.7\textwidth]{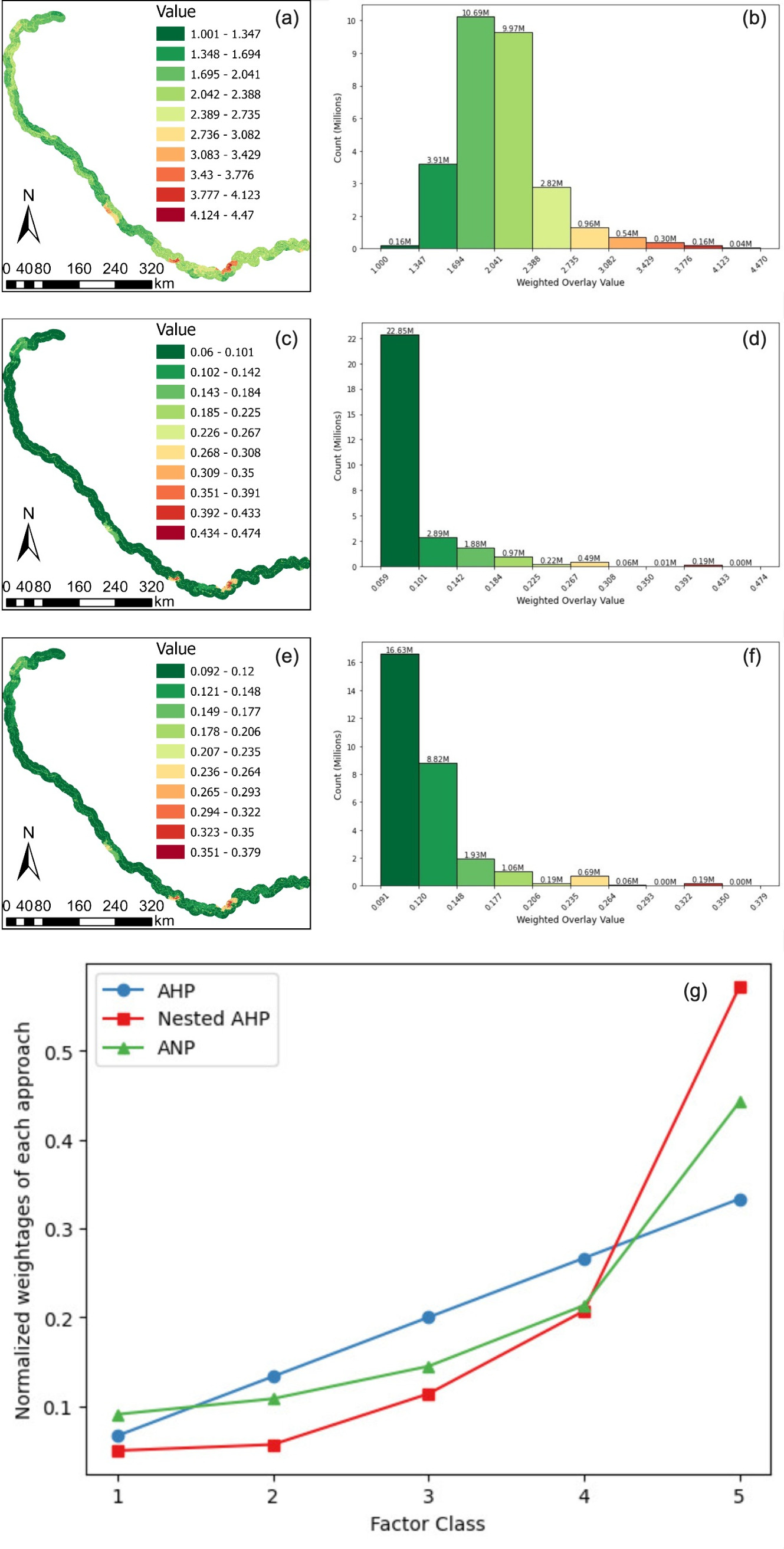}    
\end{center}
\caption{{\bf AHP, Nested AHP, ANP, and Comparison: Ganga Vulnerability Map \& Histogram} These figures show the vulnerability scores for AHP (a-b), Nested AHP (c-d) and ANP (e-f) for the scope of our analysis; the lower values pertained to the least vulnerable areas while the ones with higher scores were the most vulnerable. Map of exact values per pixel location (a,c,e) and histogram of binned ranges of values (b,d,f). The graph in (g) shows the normalized amount each factor class 1, 2, 3, 4, 5 affects the overall weighted overlay for AHP, Nested AHP and ANP, where blue (circles) represents AHP, red (squares) represents Nested AHP and green (triangles) represents ANP.}
\label{fig3}
\end{figure}

Given our pairwise matrix, we obtained the importance weights of parameters as 0.408, 0.268, 0.138, 0.100, 0.058, 0.028 for Population Density (PD), LULC, Rainfall (RAIN), Drainage Density (DD), Slope, Land Surface Temperature (LST) respectively. We multiplied the pixel values for each parameter with their weights to get our final output layer, which represented the vulnerability map of the River Ganga (Fig~\ref{fig3}).

AHP vulnerability scores ranged from a minimum of 1.001 to a maximum of 4.470 with 83.7\% of the area being categorized as having extremely low or low vulnerability and 3.5\% of the area is highly or extremely high vulnerability (Fig~\ref{fig3}(b)). Tier 2 cities (i.e., cities with a population between 50,000 and 99,999 people) on the banks of river Ganga such as Varanasi, Prayagraj/Allahabad, and Kanpur were clearly highlighted with high and extremely high AHP vulnerability scores (Fig~\ref{fig3}(a)). Indeed, urban settings that had population densities \(\geq\) 1,100 people km\textsuperscript{2} typically had vulnerability scores \(\geq\) 2.736. Alternatively, the areas with vulnerability scores \(\leq\) 1.694 were characterized by forest or cropland regions with low human population densities. The annual rainfall in the high elevation regions of the river Ganga generally increased the vulnerability of those regions.

\subsection*{Nested AHP and ANP}

We used Google Earth Engine to multiply the subweights based on the bucketed value for each factor, and multiplied it with our original AHP weights to get our final vulnerability scores for the entire Ganga buffer. The new weights for the six criteria were 0.44264, 0.21335, 0.14490, 0.10835, 0.09076 for each factor class 1, 2, 3, 4, 5. 

The distribution of vulnerability scores produced with both Nested AHP and ANP were markedly different from the distribution produced with AHP (Fig~\ref{fig3}(c-f)). Specifically, Nested AHP resulted in 96.7\% of the area scored as extremely low and low vulnerability, which was skewed with a right-tail range that extended 1.8-fold further than AHP. Similarly, ANP exhibited a right-tailed distribution with low vulnerability scores. Despite that, the Tier 2 cities had high vulnerability scores for both ANP and Nested AHP, except for Kanpur, which saw a major reduction in the vulnerability scores. These differences were due to the nonlinear response of factor weights to vulnerability rankings in both methods (Fig~\ref{fig3}(g)). Thus, these methods resulted in overall focused, yet dampened, vulnerability-score maps compared to AHP. Despite the additional insights these analyses revealed, the fundamental issue of the extent to which the defined AHP factors adequately encompass the vulnerability space remained.

For ANP, the initial values of the “relative importance of subcriteria with respect to criteria (W32)” submatrix were collected from the Nested AHP analysis. To test whether the nonlinearity of this submatrix significantly affected the ANP results, we changed the matrix values from Nested AHP to original AHP. There were only minor changes from the initial ANP results (\nameref{S4_Fig}), suggesting that the nonlinear results of ANP were independent of the submatrix values.

\subsection*{1-N AHP}

\begin{figure}[h! tbp]
\begin{center}
    \includegraphics[width=0.7\textwidth]{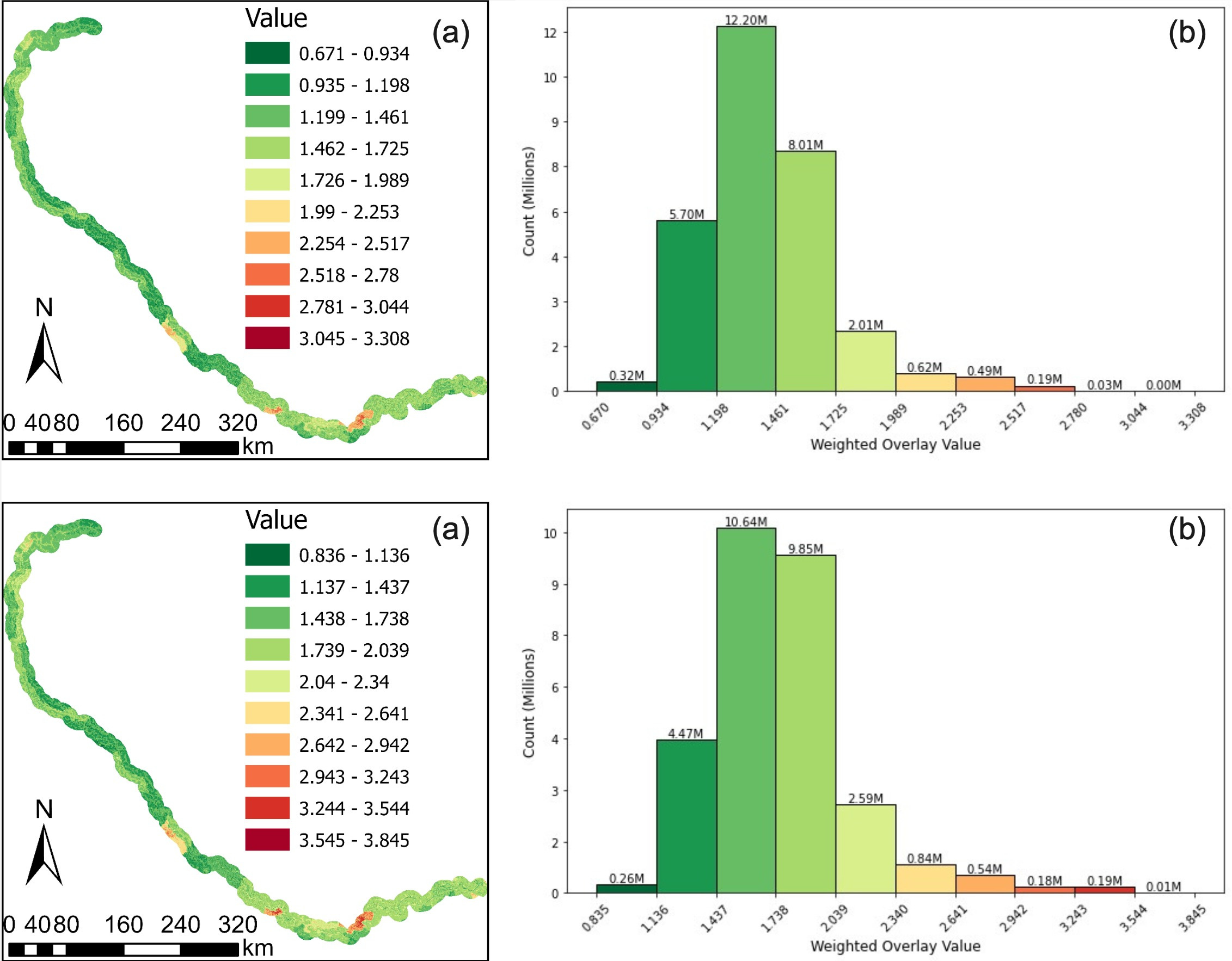}    
\end{center}
\caption{{\bf 1 - N AHP Worst case (N = 0.33) and Average case (N = 0.165): Ganga Vulnerability Map \& Histogram} These figures show the vulnerability scores for 1-N AHP when we consider the worst case (N=0.33) (a-b) and average case (N=0.165) (c-d); the lower values pertained to the least vulnerable areas while the ones with higher scores were the most vulnerable. Map of exact values per pixel location (a,c) and histogram of binned ranges of values (b,d).}
\label{fig4}
\end{figure}

When we ran 1-N AHP for the worst case, N = 0.33, we attributed more weightage to the unknown acute and chronic factors than for the average case, N = 0.165. However, due to the low probability of the presence of an acute or chronic factor, the vulnerability scores were lower for the worst case than the average case (Fig~\ref{fig4}). Even when we considered the distribution of the result values, we saw that the results for the worst case were more right skewed than the average case, which was more right skewed than the original AHP. From this we can gather that given our assumptions, there was no negative impact of considering unknown factors in our AHP analysis. However, 1-N would positively be able to capture the robustness of AHP results in other contexts, or if the assumptions were altered, and since our acute and chronic layers were purely random across the Ganga buffer, the change in vulnerability values were correlated to the values of the initial AHP results, and not to any specific remote sensing or other GIS data layers in the analysis.

\subsection*{Sensitivity Analysis with Fuzzy AHP and Fuzzy 1-N AHP}

Out of 100,000 Fuzzy AHP simulations, there were 4,433 rank reversals and no simulation had any second-order rank reversal (cases with greater than 1 index being flipped e.g. Population Density and Rainfall) at 95\% fuzziness. This implied that there was a 4.43\% chance of rank reversals at 95\% fuzziness. Of these reversals, the vast majority occurred between the factors Rainfall and Drainage Density (Fig~\ref{fig6}), which can be attributed to the low percentage difference between their initial values. 

\begin{figure}[h! tbp]
\begin{center}
    \includegraphics[width=0.7\textwidth]{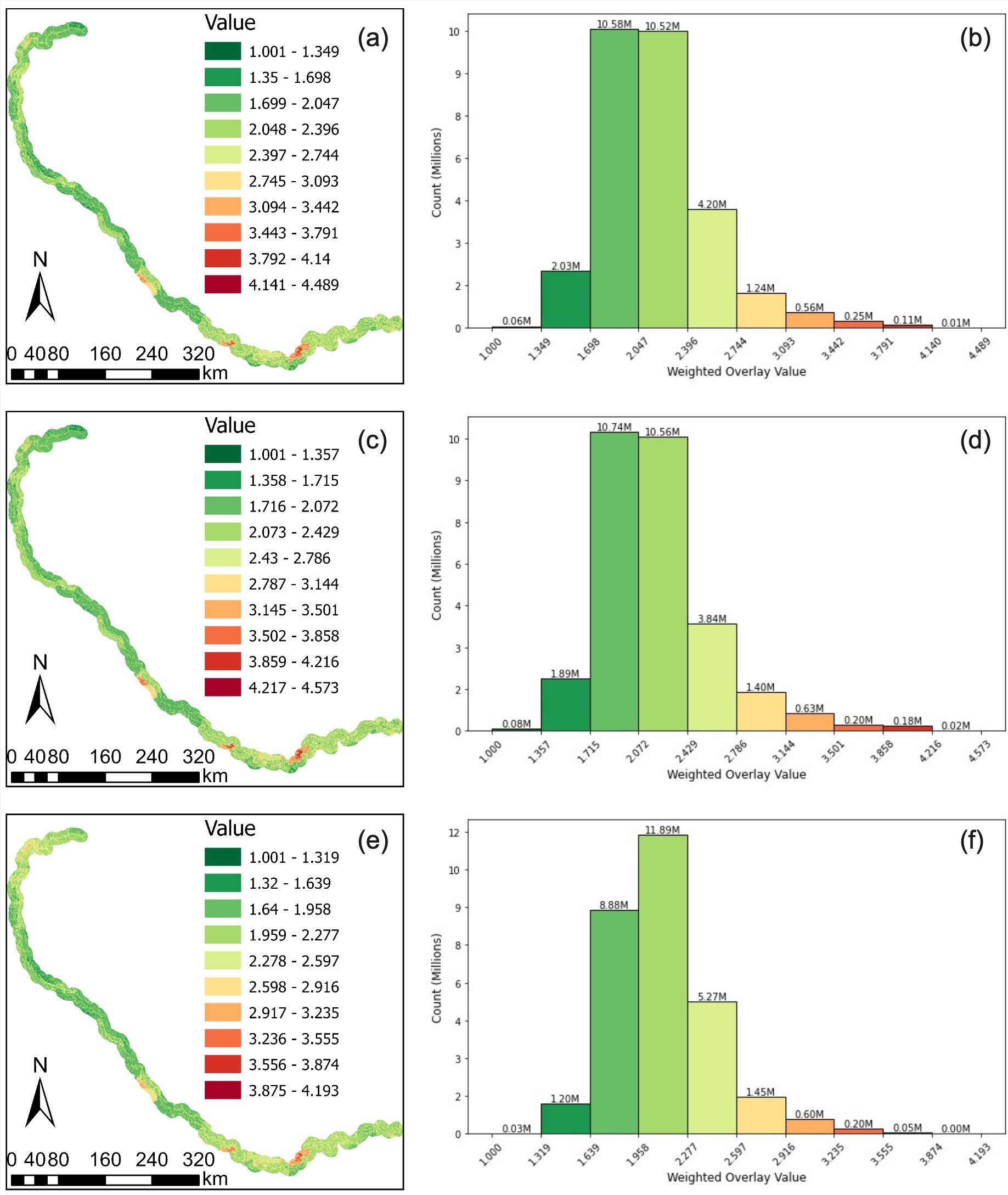}    
\end{center}
\caption{{\bf Fuzzy AHP cases (Mean, Case 1, and Case 2): Ganga Vulnerability Map \& Histogram} These figures show the  vulnerability scores for Fuzzy AHP’s Mean case (a-b), Case 1 (c-d) and Case 2 (e-f) for the scope of our analysis; the lower values pertained to the least vulnerable areas while the ones with higher scores were the most vulnerable. Map of exact values per pixel location (a,c,e) and histogram of binned ranges of values (b,d,f).}
\label{fig5}
\end{figure}

\begin{figure}[h! tbp]
\begin{center}
    \includegraphics[width=0.7\textwidth]{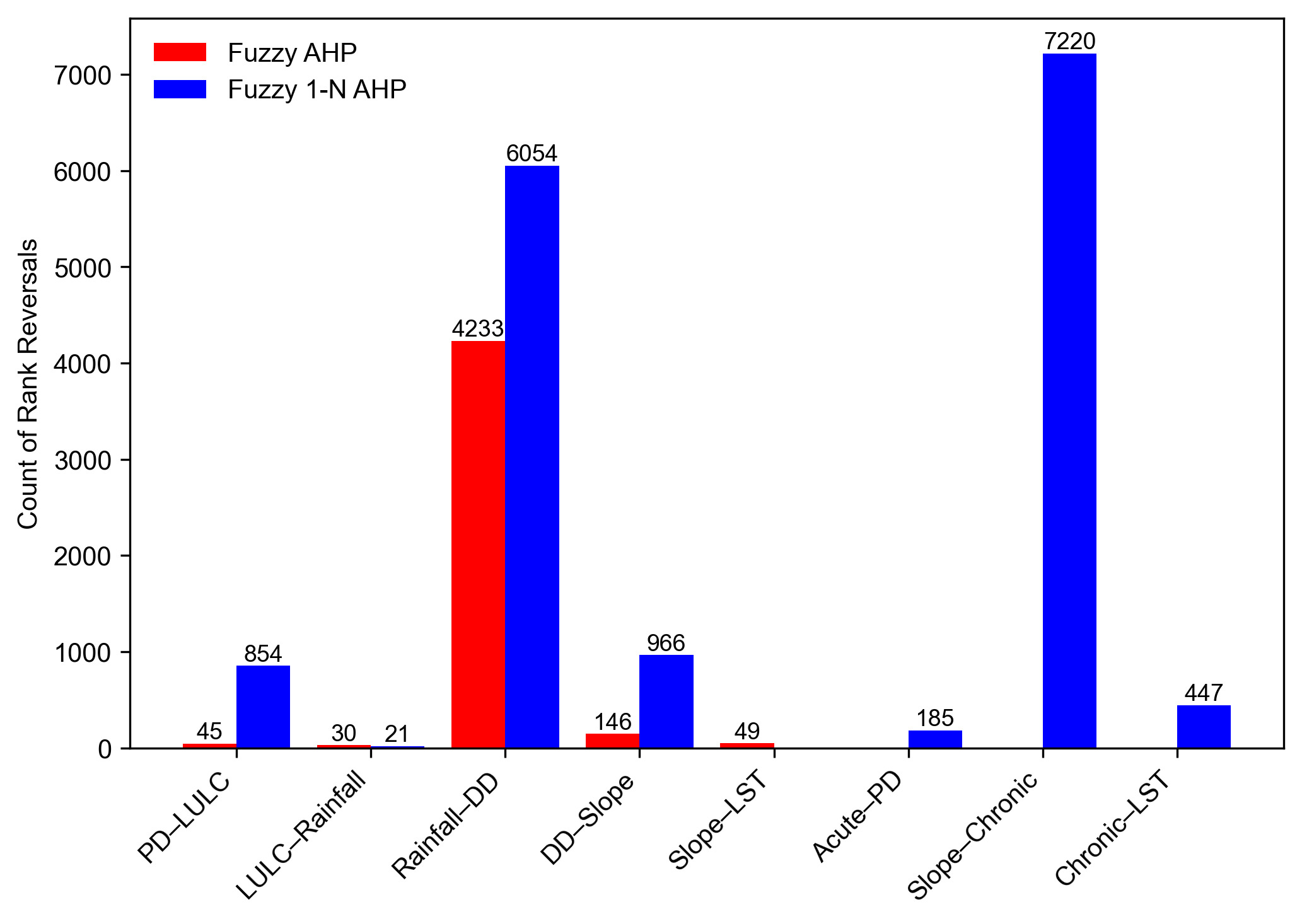}    
\end{center}
\caption{{\bf Count of First-order Rank Reversals for Fuzzy AHP and Fuzzy 1-N AHP out of 100,000 simulations} Red bars represent the count of first-order rank reversals for Fuzzy AHP, and blue bars represent the count of first-order rank reversals for Fuzzy 1-N AHP.}
\label{fig6}
\end{figure}

The simulations with rank reversals were tested to determine their potential impact on vulnerability scores. This was done by running AHP on a sample of 500 from the 4,433 cases where a rank reversal occurred, and averaging all the output layers to return the Mean Fuzzy AHP layer (Fig~\ref{fig5}(a-b)). Of these, areas that experienced changes in vulnerability due to rank reversals had high Drainage Density, or were areas that tended to receive large amounts of precipitation.

We also tested the change in vulnerability for two cases with the biggest rank reversal (i.e., the worst case). There can be two ways with the biggest rank reversals:
\begin{itemize}
    \item \textbf{Case 1:} When there was a rank reversal between the two factors that have the highest weightage (i.e., Population Density and LULC, with the highest delta between the weights).
    \item \textbf{Case 2:} When the difference between the fuzzy AHP results for a simulation had the highest delta with the values of the original AHP factor weights.   
\end{itemize}

For both the cases (Fig~\ref{fig5}(c-f)), we saw the distribution skewed to the left as compared to the original AHP, and the vulnerability increased, which was more prominent for Case 2 (Fig~\ref{fig5}(e-f)). Due to the high weightage of the parameter, there was an increase in vulnerability scores in the areas with high population density.

For Fuzzy 1-N AHP, we combined our approaches for 1-N AHP Worst case, and Mean Fuzzy AHP. There were 15,413 rank reversals out of 100,000 AHP simulations, suggesting a 15.4\% chance of rank reversals. Most of the rank reversals were between the factors Slope and Chronic, followed by Rainfall and DD, similar to Fuzzy AHP (Fig~\ref{fig6}). There were around 200 second-order rank reversals for Fuzzy 1-N AHP, between the factors Slope and LST (117), DD and Chronic (77), and Rainfall and Slope (5). A major contributor to the rank reversals was the closeness of the original AHP weights, leading to a higher likelihood of rank reversals when fuzziness is incorporated.

\subsection*{Comparison of analysis scores}

To compare different approaches and test areas where the predicted vulnerability changes, we normalized the different output layers by standardizing them to follow Normal distributions with a Mean of 0 and Standard Deviation of 1, i.e., Standard Normal Distribution. This enabled us to compare different approaches that had different ranges. Then we subtracted each output layer from the original AHP results, i.e., AHP Vulnerability Score - Variant Vulnerability Score, to get our final distribution of difference layers. While computing these layers, the positive values showed that as per AHP, the areas were identified as more vulnerable than the variant map did, and the vice versa in case of negative values.

We compared the results of every approach (except Fuzzy 1-N AHP) with the original AHP results, using AHP as the baseline for checking the robustness of our approaches. Fuzzy 1-N AHP was excluded from the comparisons analysis because it provided little additional information (\nameref{S7_Text}). We wanted to infer three things from the distribution of difference layers, 1) Which pixel values changed relative to AHP, 2) How much did they change, 3) Were the changed values higher or lower than AHP?

Also, since the effect of Population Density on vulnerability was high, we conducted the same difference test to assess whether the Population Density layer by itself could suffice in assessing the vulnerability of the basin. This tells us how much of the variability of the vulnerability score can be attributed solely to Population Density.

To visualize and compare the distributions of differences of the AHP layer with other approaches, we created 3 different map types (\nameref{S5_Fig}). We created Map Type 1 by stretching the colors between the overall minimum and maximum of all the different approaches, to compare the approaches among themselves. For Map Type 2 we stretched the colors between the local minimum and maximum value of each approach, to make outliers and all instances with high variability (\(\geq\) 2 standard deviations (SD)) for each individual map stand out. And for Map Type 3, we stretched the colors between the local 5 percentile and 95 percentile for each approach, to prevent the values \(\geq\) 2 SD from washing-out the differences in the map.

\subsubsection*{Difference between Nested AHP, ANP and AHP} Observing Nested AHP and ANP, we saw a similar pattern of differences with AHP, owing to the non-linear distribution of both the results. While the AHP predicted the vulnerability to be high in Kanpur, ANP and Nested AHP suggest that it would be lower. And the vice versa is seen to be true for Prayagraj/Allahabad and Varanasi, where AHP predicted the vulnerability to be lower than the results of ANP and Nested AHP. To test whether the new layers add additional information over the AHP layer, we performed a two-way ANOVA test with the parameters quintile and model type (AHP, ANP, Nested AHP) (\nameref{S5_Table}). For this, we took out the quintile values for pixels in each layer, and sampled 1000 random points from each AHP quintile. The p-values for quintiles, and the combined effect of quintile and model type on the vulnerability scores were <0.0001, but the p-value for models only was 0.6978. This is likely due to the lack of splitting the values to their respective quintiles. When we performed a Tukey post-hoc test for each quintile, we found that there was a significant difference between the variants within each quintile. This confirms a statistically significant difference among Nested AHP, ANP, and AHP quintiles, and highlights the methodological gains offered by combining Nested AHP and ANP with classic AHP.

\subsubsection*{Difference between 1-N AHP and AHP} Despite the different visualization approaches, there were no major visual differences for the vulnerability scores of 1-N AHP with AHP. Given that the distributions within the first and fifth quintile are non-normal, we performed the non-parametric Mann-Whitney U test, and for the other three quintiles, we ran the paired t-test. For all the tests comparing AHP and 1-N AHP (Average and Worst case) within each quintile (except for quintile 3, where  p=0.8955 for AHP and 1-N AHP Average case, and p=0.3911 for AHP and 1-N AHP Worst case) p-values were less than 0.001, suggesting that there was a significant difference between the AHP and 1-N AHP layers. The lack of statistical difference in quintile 3 can be explained by the probabilistic layers of acute and chronic error, and the fact that the factors other than acute and chronic are strongly correlated with the AHP variables.

\subsubsection*{Difference between Fuzzy AHP cases (Mean, Case 1, and Case 2) and AHP} In the Mean Fuzzy AHP Case, we sampled some of the simulations, and most of them had a rank reversal between rainfall and drainage density, causing those layers to reduce the robustness of the results. In Case 1 and Case 2, both approaches were in agreement with each other across the latter half of the Ganga buffer, but we see that in the high-altitude areas they were tending in opposite directions. In Case 1 the AHP suggested a higher vulnerability than Fuzzy AHP, primarily affected by Population Density, and in Case 2, Fuzzy AHP returned a higher value for vulnerability scores, owing to the Annual Rainfall variable. We performed a two-way ANOVA test with the parameters quintile and model type (AHP, Mean Fuzzy AHP, Fuzzy AHP Case 1, Fuzzy AHP Case 2) (\nameref{S6_Table}). The p-values for quintiles, model type, and their combined effect on the vulnerability scores were all less than 0.0001. When we performed a Tukey post-hoc test for each quintile, we found that there was no significant difference between the following variant pairs within the quintiles: AHP and Mean Fuzzy AHP (Quintile 1, 3), AHP and Fuzzy AHP Case 1 (Quintile 2, 3), Mean Fuzzy AHP and Fuzzy AHP Case 1 (Quintile 2, 3, 4). This confirms a statistically significant difference among AHP and Fuzzy AHP approaches, and highlights the methodological gains offered by Fuzzy AHP.

\subsubsection*{Difference between Population density and AHP} The differences between the standardized values of AHP and the Population density layers were the highest among all approaches, and there were many areas where there is a non-zero value for this layer (\nameref{S5_Fig} 8.a). This indicates that AHP is amply influenced by the other parameters and proves the effectiveness of the AHP approaches for creating a good composite variable. This can further be validated by the t-tests and Mann-Whitney U tests, where within each quintile (except for quintile 3, where p=0.4026) we get a p-value less than 0.001, suggesting that there was a significant difference between the AHP and Population density layers.

\subsubsection*{Graphical Representation of Variances} 
\begin{figure}[h! tbp]
\begin{center}
    \includegraphics[width=0.7\textwidth]{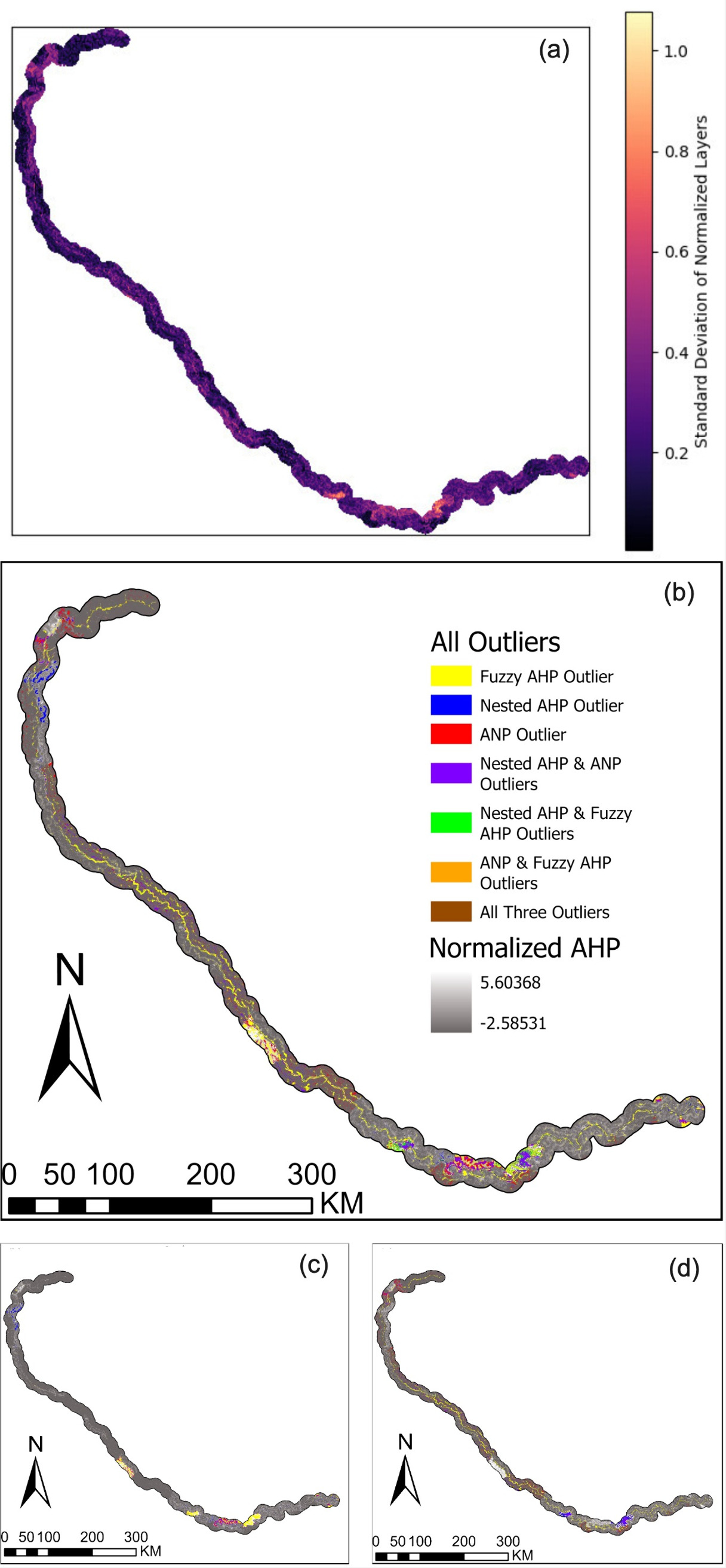}    
\end{center}
\caption{{\bf Standard Deviation of Normalized and Composite Layers} Standard deviations of Normalized layers (a) for the approaches AHP, ANP, Nested AHP, where darker values suggest lower standard deviation, and the lighter values suggest high-standard deviations. The composite layer (b) that can be used to describe the vulnerability of the river Ganga. We assign different colors for pixels representing \(\geq\) 2 SD of each variant, and to showcase the positive and negative instances, we split them into two charts (c) and (d).}
\label{fig7}
\end{figure}

To compare the predictions for multiple variants at once, we calculated the standard deviation between three approaches: AHP, ANP, Nested AHP for each pixel and visualized them for our region of interest (Fig~\ref{fig7}(a)). We noticed that the deviations were high for the tier-two cities Prayagraj/Allahabad and Varanasi. Varanasi had the highest vulnerability in terms of AHP, but the other variants of AHP brought more emphasis to Prayagraj/Allahabad as well. There were also some high deviations in the high-altitude regions near Rishikesh - a high-population city in Uttarakhand, suggesting disagreements between the approaches. 

\subsection*{Composite Layer for all Approaches}

\begin{figure}[h! tbp]
\begin{center}
    \includegraphics[width=0.7\textwidth]{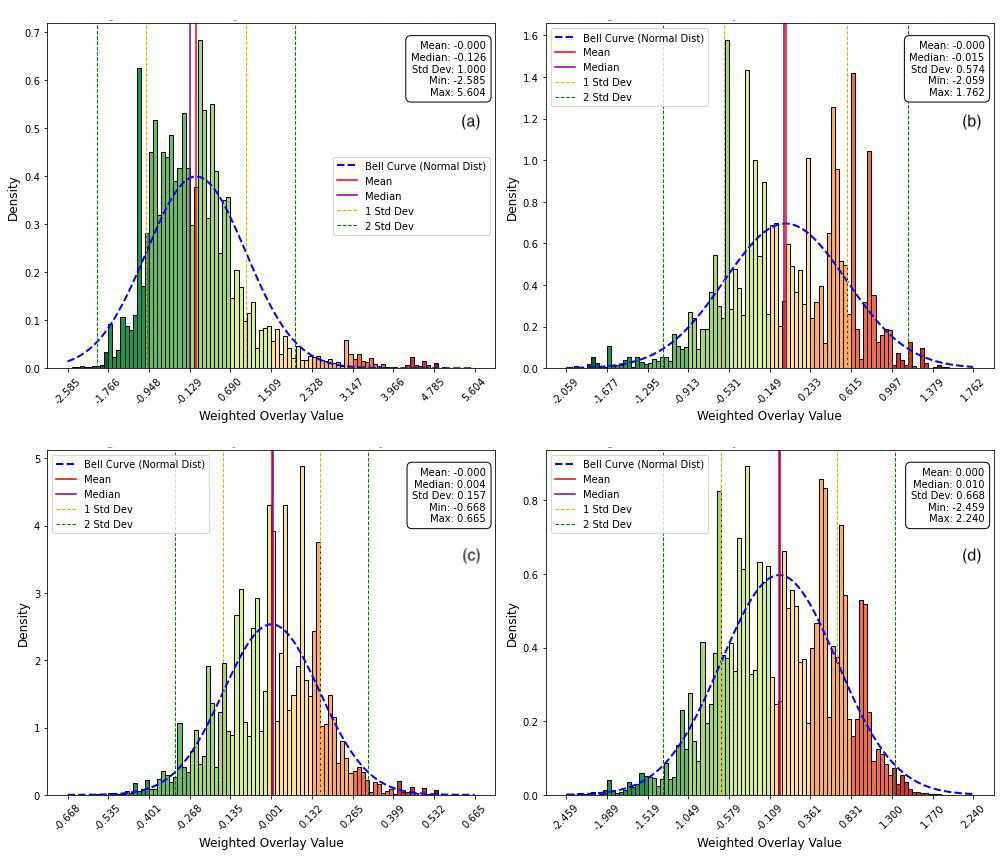}    
\end{center}
\caption{{\bf Normalized AHP \& its Difference with Normalized AHP Variants} Here, (a) shows the Normalized AHP layer with its standard deviation, while the rest of the figures show the distribution of the differences between Normalized AHP and its corresponding AHP Variants: ANP, Mean Fuzzy AHP, and Nested AHP. (b) is Normalized AHP minus Normalized ANP. (c) is Normalized AHP minus Normalized Mean Fuzzy AHP. (d) is Normalized AHP minus Normalized Nested AHP.}
\label{fig8}
\end{figure}

To create a composite layer to capture the effect of each variant of AHP across the river buffer, we took a normalized base layer for AHP, and overlaid the outliers (\(\geq\) 2 SD) for each of the difference layers for Mean Fuzzy AHP, Nested AHP, and ANP (Fig~\ref{fig7}(b-d)).  To understand the distribution of the AHP base layer, and the distribution of the difference layers for Mean Fuzzy AHP, Nested AHP, and ANP, we also created the histograms of each layer (Fig~\ref{fig8}). In the composite layer, the primary colors (red, blue, yellow) were used to depict the outliers for the individual layers, and secondary colors (orange, green, purple) were used when the effects of two AHP variants were combined. Given that the outliers could be both positive and negative, we also created separate maps for the two. If AHP is greater than the variant, the outliers are positive, and vice versa. These composite layers can be used as a way to depict a single variable for river Ganga’s vulnerability with the effect of all AHP variants. Through these composite layers, we are illustrating the effect of AHP relative to the unique effects of the variants wherever they disagree with AHP.

\begin{figure}[h! tbp]
\begin{center}
    \includegraphics[width=0.7\textwidth]{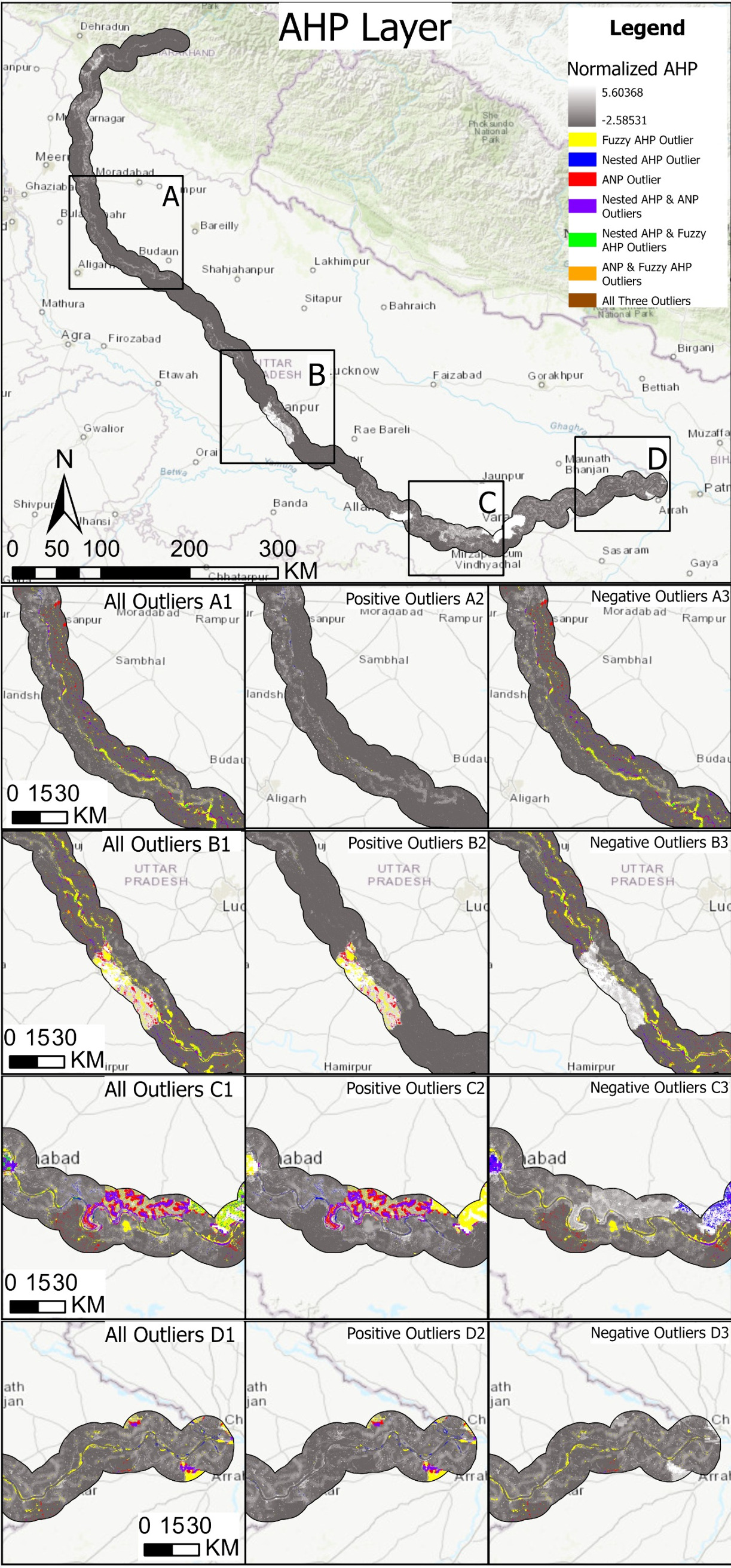}    
\end{center}
\caption{{\bf Focused Composite Layers} Subsection analysis of the composite variable for four subsections A, B, C, D, with different panels numbered 1 (all instances of \(\geq\) 2 SD), 2 (positive instances), and 3 (negative instances). To cover a range of distributions, we considered two rural (A, D), and two urban (B, C) subsections. Out of these, Subsection B has the city Kanpur in them, while subsection C is near Prayagraj/Allahabad and Varanasi, highly populated areas along the Ganga.}
\label{fig9}
\end{figure}

We also selected four important subsections of the river buffer where the composite variable was effective in capturing the effects of each AHP variation (Mean Fuzzy AHP, Nested AHP, and ANP). There were some small areas near the river body which had high variability between AHP and all three AHP variants. We can see the effect of the combination of two variants in Fig~\ref{fig9} subsection C, where there was a lot of variability for both Nested AHP + ANP (violet) and Nested AHP + Mean Fuzzy AHP (green) variants. The composite vulnerability threshold was set to \(\geq\) 2 SD so that most inter-model variance was silenced, yet differences that exceeded 95\% were highlighted. These differences distinguish areas of note such as highlighting particularly vulnerable locations (ANP and Nested AHP); identifying instances where network effects are more important than differences in factor class (ANP) and vice versa (Nested AHP); and cataloging when AHP scores are sensitive to factor weights (Fuzzy AHP). These composite highlights may be used by policymakers and project developers to guide specific Ganga River management interventions (e.g., infrastructure upgrades, conservation zoning).  For instance, the violet areas surrounded by red highlight areas near Varanasi (Fig~\ref{fig9} subsection C) may be prioritized for Ganga rejuvenation efforts.

\section*{Discussion}

In this paper, we aimed to assess the pollution vulnerability levels of the river Ganga using remote sensing datasets. We lacked ground truth data for pollution, making supervised learning algorithms unfeasible, and making it difficult to validate the metric. While potential proxy datasets (e.g., river water quality indices, remote sensing-based chlorophyll or turbidity measures) can help validate model outputs, they are limited both spatially and temporally. Furthermore, access to the data may be restricted, or they are part of reports and not easily extracted. For instance, a report on Ganga biological health was published in 2017, yet the data were collected in 2014-2016 from 93 stations along the entire length of Ganga and identifying the sites required cross referencing a station lat/long table with station names and data parsed into different tables devoted to specific river sections \cite{32}. Similarly, religious festivals, like Magh Mela (every January and February) and Kumbh Mela (every 12-13 years), may also dramatically impact pollution loads (Tiwari et al. \cite{33}). Thus, we utilized the widely used AHP decision making method and publicly available GIS data layers in order to generate a spatial numeric dataset reflecting the pollution vulnerability of the river Ganga. Gaining confidence around our results is critical, as decisions regarding the river’s ecological health depend on a reliable assessment of vulnerability levels and  ground truth data (e.g., water samples) should be incorporated into future sampling schemes guided by vulnerability scores, especially in urban-adjacent hotspots. These future efforts may be used for model validation in addition to project development and act as a self reinforcing vulnerability assessment that supports local and national monitoring efforts by the Central Pollution Control Board and Namami Gange.

Despite the advantages of AHP, it has multiple shortcomings (e.g., it may be influenced by potentially biased expert opinion, it is unable to handle factor interdependencies, and it is limited to a defined factor decision space). So we explored existing alternative methods, such as ANP and Fuzzy AHP, and we designed novel variants of AHP, such as Nested AHP, 1-N AHP, and a hybrid Fuzzy 1-N AHP. Each of these approaches offered unique advantages over the original AHP approach, and contributed towards the robustness of our analysis by addressing different shortcomings of the AHP method.

For instance, AHP functions under the assumption that the criteria are independent of each other (\nameref{S1_Table}, 5.2.8), which is fairly uncommon to achieve in real world scenarios, especially in ecological settings. For this reason, we used ANP to capture the criteria dependencies, and the results of ANP were accommodated into our proposed solution. Another shortcoming being addressed by our approach is that we tested the uncertainty in the decision space (\nameref{S1_Table}, 5.2.24) by checking the likelihood of rank reversals using Fuzzy AHP and Fuzzy 1-N AHP, thus ensuring that our solution is robust to uncertainties in the decision space. Within Fuzzy AHP, we introduced another novel methodology which involved using Monte Carlo simulations to get randomized values of fuzziness instead of relying on the domain experts. This way we managed to quickly run a variety of scenarios of AHP while excluding the need of domain experts. Lastly, the “external” influence on decision (\nameref{S1_Table}, 5.2.26) is accommodated by expanding to the breadth of all possible factors that might influence the decision space using 1-N AHP.

To conclude, we incorporated the variability of these different approaches into a new robust composite variable, one which overcomes the shortcomings of AHP. We can use this ensemble variable as an adaptable and reliable decision making tool to plan remediation methods for the river Ganga. 

Although this study addressed many of AHP’s known shortcomings, there were some limitations that were not addressed. Of note, the need for multiple technical experts to address the broad range of factors when assigning Saaty rankings is a valid concern. 

While we are an interdisciplinary group, we are not subject matter experts across all of the fields represented by the AHP input factors. To address this issue, an overall quality metric of the Saaty rankings may mitigate this issue based on the level of expertise and available data (e.g., a quality metric might consist of four levels, ranging from novice opinion to the consensus of subject matter experts). Alternatively, a Saaty scale for quality could range from -1 to +1, where misinformation/disinformation could be ranked with -1; poorly defined/constrained data with little pertinent value could be assigned a value near zero; and data based on world class consensus of experts could be ranked as +1. A gradient may be applied to distinguish levels of confidence similar to IPCC consensus rankings \cite{35}. Furthermore, lack of certainty about the importance of alternative criteria and biases during ranking may be addressed by running different scenarios, as long as the decision space is fundamentally partitioned based on an overarching difference due to a given circumstance, and robust data and guidance is provided when rankings are calculated. In addition, vulnerability scores were dependent on remote sensing data resolution, which varied by data layer (e.g., population density is reported at 1 km\textsuperscript{2} scale and digital elevation at 30 m). For instance, low resolution population density values homogenize areas under a factor class, which may result in greater influence of population along sparsely populated edges of urban areas while muting the local impact of small densely populated villages in rural areas. Data resolution also varied temporally (e.g., population data were based on decadal survey results and rainfall data were annual amounts), which needs to be considered during planning and decision making. Moreover, vulnerability scores were context sensitive such that decision spaces may vary depending on the influence, conditions, and/or regulatory as well as technical mitigation potential in different biomes, nations, and jurisdictions to alleviate sources of vulnerability. Furthermore, although publicly available remote sensing data were used in this study, which gives greater opportunity for replication and use of the approach in other locations such as the Nile and Mekong Basins when weights are adjusted for local factor importance, the incorporation of limited-accesss, higher-resolution remote sensing datasets to improve accuracy is possible and would benefit policymakers. Finally, dependencies and correlations between or among factors may not be well understood. While this is true, the process is iterative such that the models can be refined as additional data are gathered, factor interactions are clarified, and scenarios are tested.

Additional data layers such as GIS-based applications that track impacts may be integrated into the data process to support and hone vulnerability assessments and thus AHP factor weights as well as improve output resolution and validation. For instance, acute factors may be known unknowns (e.g., mining locations) and unknown/ill defined unknowns (e.g., location and breadth of endangered species habitat). Similarly, potential widespread impacts such as loss of biodiversity and fragmentation of biodiversity richness are chronic factors that were excluded from our AHP vulnerability decision space but may be included in future efforts. Indeed, the approach of sand mining and waste disposal tracked by Bayazidy et al. \cite{2} may be used to identify specific instances of these and other acute factors that require precise and timely identification. In addition, coupling biodiversity and the conservation imperative analysis of Dinerstein et al. \cite{5} with our vulnerability assessment can further prioritize areas for conservation that were scored as low biodiversity and extremely low vulnerability. Moreover, emerging factors may be added as data are collected and the breadth and depth of the decision space grow. For instance, the environmental spread of antibiotic resistance in freshwater is tracked with molecular genetic techniques~\citep{29,11}, and coupling these data with their spatial locations can add antibiotic resistance as a vulnerability factor. Finally, high-resolution data from UAV such as that produced by Tripathi et al. \cite{27} during their effort to combine remote sensing with drone data while cataloging riparian vegetation along the Ganga can augment and further refine LULC information to better refine vulnerability scores.

Ultimately, networks of models may be brought together to enrich landscape digital twins for design, decision making, and monitoring of real world features of landscapes. These twins become digital representations of natural and built features in landscapes. For river systems, digital twins may consist of riverscapes, the concept that integrates longitudinal characteristics of rivers with their land features~\citep{31}.  Indeed, the basis for such twins already exists in regions prone to flooding. For instance, Nested AHP was used to predict five categories of flood susceptibility with nine factors based on remote sensing data in addition to soil texture, geomorphology, and geology~\citep{4}. Furthermore, AHP with nesting of four factors (land use, percent green space, per capita sewer length, and slope) was used to produce a flood vulnerability index in a section of Hanoi, Vietnam~\citep{10}, and a flood susceptibility map for all of Hanoi was made with a 9-factor AHP that was ranked by subject matter experts~\citep{14}. 

Although vulnerability to pollution is the focus of this study, human socio-economic factors (e.g., public health and poverty data, access to clean water and sanitation services, industrial impacts) need to be assessed when making holistic vulnerability assessments. Indeed, AHP models that include cultural, socioeconomic and environmental criteria provide for wider assessments of the human condition. Recently, 20 factors representing these criteria as well as security and service functions were placed in a hierarchical structure and assessed through a survey of subject matter experts to realize a quality of life score~\citep{28}. Of note, they detected the inverse of our vulnerability scores in that quality of life tended to decrease away from city centers and the services they provide whereas cities consistently were sources of high vulnerability to the river Ganga. Thus, the perspective of the modeling scenario must be understood and integrated into the larger decision space of balance between human endeavors and nature. Moreover, a rich diversity of alternative perspectives can drive sustainable development so that factors that improve quality of life can be implemented with policies that recognize and mitigate their accompanying sources of vulnerabilities. Importantly, models may be used to inform and guide policy as jurisdictions aim to meet just, equitable, and sustainable development in parallel with the ten principles for ecosystem restoration developed by the United Nations \cite{34}.

Ganga Mitra of the National Mission for Clean Ganga is focused on the environmental governance and rejuvenation of the river Ganga. The vulnerability assessment presented here may be implemented to meet its core mission “to conserve water \& natural resources for livelihoods \& sustain the life in Ganga Basin”. Indeed, the vulnerability scores provide an unparalleled system overview that may be harnessed to identify, prioritize, and target on-the-ground testing, monitoring, and project development for Ganga riverscape rejuvenation and conservation. Moreover, integrated multidimensional models and digital twins have the potential to provide policy makers practical applications for water management and restoration efforts. For instance, an actionable roadmap for the Ganga Basin is as follows: conduct high resolution LiDAR-based survey of the Ganga riverscape, combine local and national industrial and socioeconomic data with social media data and quality of life metrics, integrate the multidimensional into a suite of indices, and model potential alternatives against desired priorities and outcomes.

\section*{Conclusion}

In this study, we assessed the vulnerability of the river Ganga to pollution along a 20 km wide corridor covering 26,609.4 km\textsuperscript{2} of riverscape. We used AHP to perform a dimensionality reduction of six remote sensing datasets to create a single variable for river vulnerability.  To mitigate structural limitations to AHP, our approach brought together a suite of methods and produced a set of comparative metrics to identify and refine vulnerability scores. Of note, urban landscape features with extremely high vulnerability scores and those at the interface with vulnerable areas were identified. These findings provide the basis to rational prioritization of pollution mitigation and a list of locations for future rejuvenation. Moreover, sites currently with low vulnerability scores may be targeted for conservation or sustainable development practices to prevent their degradation. Our vulnerability score, together with other metrics such as a quality of life score~\citep{28}, can enrich digital twins beyond simplistic animations and reveal their underlying environmental, social, and economic characteristics.

\section*{Author contributions}

\textbf{Conceptualization:} Sarthak Arora, James C. Smoot, Nikhil Raj Deep, Claire Gorman Hanly, Anthony Acciavatti.\\
\textbf{Data curation:} Sarthak Arora, Michael Warner, Ariel Chamberlain.\\
\textbf{Formal analysis:} Sarthak Arora, Ariel Chamberlain.\\
\textbf{Investigation:} Sarthak Arora, Nikhil Raj Deep.\\
\textbf{Methodology:} Sarthak Arora, Nikhil Raj Deep, Claire Gorman Hanly. \\
\textbf{Project administration:} James C. Smoot, Anthony Acciavatti.\\
\textbf{Supervision:} James C. Smoot, Anthony Acciavatti.\\
\textbf{Visualization:} Sarthak Arora, Michael Warner, Ariel Chamberlain.\\
\textbf{Writing – original draft:} Sarthak Arora, Michael Warner, James C. Smoot.\\
\textbf{Writing – review \& editing:} Sarthak Arora, Michael Warner, Ariel Chamberlain, James C. Smoot, Nikhil Raj Deep, Claire Gorman Hanly, Anthony Acciavatti.\\


\section*{Acknowledgments}
The authors thank Markley Boyer and Sarah Bergmann for thoughtful discussions throughout the project. We also thank Aaron Hirsh for organizing Collaborative Earth. Furthermore, we thank Dr. Mohamed A. E. AbdelRahman, Division of Environmental Studies and Land Use, National Authority for Remote Sensing and Space Sciences (NARSS), Cairo, Egypt, and five anonymous reviewers for providing comments that greatly improved this paper.

\nolinenumbers

\section*{Supporting information}

\paragraph*{S1 Table.}
\label{S1_Table}
{\bf Shortcomings of AHP analysis rubric.}
\begin{adjustwidth}{-2.25in}{0in}
\centering
\renewcommand{\arraystretch}{1.2} 
\begin{tabular}{@{}p{1.8cm}p{4.5cm}p{1.2cm}p{4.5cm}p{2.5cm}@{}}
\toprule
\textbf{Critique No.} &	\textbf{Description} &	\textbf{Relevant (Y/N)} &	\textbf{Comments} &	\textbf{Suggested Alternatives} \\
\midrule
5.2.1 &	The Pair-Wise Method and Its Application in AHP &	N &	 &	 \\
5.2.2 &	The Pair-Wise Method in AHP Constructs Artificial Relationships &	Y &	Test the robustness of the analysis, and define its uncertainty as well as consistency. &	Fuzzy AHP \\
5.2.3 &	Criteria Preferences Must Consider Alternatives &	N &	 &	 \\
5.2.4 &	The Ambiguity of Pair-Wise Comparisons in AHP &	N &	 &	 \\
5.2.5 &	Modelling Scenarios &	N &	. &	 \\
5.2.6 &	Lack of Rational Answers from DMs &	N &	 &	 \\
5.2.7 &	AHP Incapacity to Solve Complex Problems &	Y &	Complex problems have trade-offs of varying importance and levels of certainty &	1-N AHP, ANP, Fuzzy AHP \\
5.2.8 &	Criteria Independency &	Y &	A testing of marginal effects is warranted to understand the strengths of interactions. &	1-N AHP, ANP \\
5.2.9 &	Quantifying Preferences &	N &	 &	 \\
5.2.10 &	Quantitative Data &	N &	 &	 \\
5.2.11 &	The DM Should Be a Multiple Technical Expert &	Y &	It’s a valid concern and surveys of experts to find a consensus (and range) of expert assessments. &	 \\
5.2.12 &	The AHP Tries to Subordinate Reality to a Theory &	Y &	Using different decision matrices for different categories of decision (e.g., cost, siting, restoration category) is reasonable. &	1-N AHP \\
5.2.13 &	Relative Importance Between Criteria Is Considered Constant &	Y &	This is the crux of \cite{S_Steele} concern with respect to relative normalization and absolute normalization which boils down to the decision space is unlikely to be completely known. &	1-N AHP \\
5.2.14 &	AHP’s Fundamental Scale Has Limits 1 and 9 &	N &	 &	 \\
5.2.15 &	A Preference and the Peculiar Meaning of Its Inverse Value &	N &	 &	 \\
5.2.16 &	Determination of Criteria Importance Must Contemplate the Alternatives &	Y &	Different scenarios should be run. If the decision space is fundamentally partitioned based on an overarching difference due to a given circumstance. &	 \\
\bottomrule
\end{tabular}
\end{adjustwidth}

\begin{adjustwidth}{-2.25in}{0in}
\centering
\renewcommand{\arraystretch}{1.2} 
\begin{tabular}{@{}p{1.8cm}p{4.5cm}p{1.2cm}p{4.5cm}p{2.5cm}@{}}
\toprule
\textbf{Critique No.} &	\textbf{Description} &	\textbf{Relevant (Y/N)} &	\textbf{Comments} &	\textbf{Suggested Alternatives} \\
\midrule
5.2.17 &	The DM Preferences Do Not Consider the Real World, They Only Exist in His Own Universe &	Y &	There may be differences with an analysis based on subjective relative importance; however, it should be minimal with robust data metrics and uncertainty measurements. &	 \\
5.2.18 &	Where Is the Logic of the DM Correcting Himself Just for the Sake of Transitivity? &	N &	 &	 \\
5.2.19 &	Normalization &	Y &	1-N AHP is a test of this. &	1-N AHP \\
5.2.20 &	The Use of the Eigenvector Method &	N &	 &	 \\
5.2.21 &	Criteria Weights &	Y &	Trade-offs may occur while subjectively assigning a rank to the relative importance of different criteria. The value/relevance is how well the decision space captures the question asked. &	 \\
5.2.22 &	A Project Is a System &	Y &	ANP may interrogate/reveal relationships &	ANP \\
5.2.23 &	Sensitivity Analysis problems &	Y &	Conventional Sensitivity Analysis is inappropriate for weights derived from subjective assessments of relative importance. Fuzzy AHP tests uncertainty rather than inconsistency. &	Fuzzy AHP \\
5.2.24 &	Rank Reversal (RR) in AHP &	Y &	Rank reversals arise when uncertainty in the decision space is larger than the consistency of rank, and it is important to test for their likelihood and the conditions that drive their occurrences. &	1-N AHP, Fuzzy AHP \\
5.2.25 &	Time Dependency in a Portfolio of Projects &	Y &	A work flow similar to the one we implemented during this study makes repeating AHP straightforward. &	 \\
5.2.26 &	This talks about “external” influence on decision &	Y &	1-N is a test of this, as an external force suggests an incomplete decision space &	1-N AHP \\
5.2.27 &	Not all problems have a hierarchy &	N &	 &	 \\
5.2.28 &	Criteria limiting other criteria &	N &	 &	 \\
5.2.29 &	Problem structuring &	N &	 &	 \\
5.2.30 &	Dependencies, correlations &	Y &	Dependencies and correlations are indirectly related to relative importance, which may be assessed with ANP. &	ANP \\
\bottomrule
\end{tabular}
\end{adjustwidth}

\paragraph*{S2 Table.}
\label{S2_Table}
{\bf Data Sources}
\begin{table}[H]
\begin{adjustwidth}{-2.25in}{0in}
\centering
\renewcommand{\arraystretch}{1.2} 
\begin{tabular}{@{}p{2.5cm}p{6.5cm}p{1.5cm}p{3cm}@{}}
\toprule
\textbf{Name} &	\textbf{Description} &	\textbf{Location\textsuperscript{a}} &	\textbf{References} \\
\midrule
Buffer Ganga river &	 &	This study &	This study \\
NASADEM &	NASA NASADEM Digital Elevation 30m (3 bands) &	\href{https://developers.google.com/earth-engine/datasets/catalog/NASA_NASADEM_HGT_001}{Link} & \cite{S_NASA} \\
WorldCover &	ESA WorldCover 10m v200 &	\href{https://developers.google.com/earth-engine/datasets/catalog/ESA_WorldCover_v200}{Link} & \cite{S_Zanaga} \\
GPWv411 &	UN-Adjusted Population Density (Gridded Population of the World Version 4.11) &	\href{https://developers.google.com/earth-engine/datasets/catalog/CIESIN_GPWv411_GPW_UNWPP-Adjusted_Population_Density}{Link} & \cite{S_CIESIN}\\
CHIRPS Daily &	Climate Hazards Group InfraRed Precipitation With Station Data (Version 2.0 Final) &	\href{https://developers.google.com/earth-engine/datasets/catalog/UCSB-CHG_CHIRPS_DAILY}{Link} & \cite{S_Funk} \\
Landsat &	USGS Landsat 8 Level 2, Collection 2, Tier 1 &	\href{https://developers.google.com/earth-engine/datasets/catalog/LANDSAT_LC08_C02_T1_L2}{Link} &	\cite{S_USGS} \\
\bottomrule
\end{tabular}
\vspace{0.5em}
\begin{minipage}{0.75\textwidth}
    \textsuperscript{a}\footnotesize Data identified as from this study is available at \href{https://drive.google.com/drive/folders/1ihsreY9NwbZyCLfNZALGslUK0BZ0o1BX?usp=sharing}{Drive Link}
    \end{minipage}
\end{adjustwidth}    
\end{table}

\paragraph*{S3 Table.}
\label{S3_Table}
{\bf AHP Pairwise and Normalized Comparison Matrix}
\begin{table}[H]
\begin{adjustwidth}{-2.25in}{0in}
\centering
\renewcommand{\arraystretch}{1.2} 
a.) \textbf{AHP pairwise comparison matrix\textsuperscript{a}}
\\
\begin{tabular}{@{}p{1.25cm}p{1.25cm}p{1.25cm}p{1.25cm}p{1.25cm}p{1.25cm}p{1.25cm}@{}}
\toprule
 \textbf{} & \textbf{PD} & \textbf{LULC} & \textbf{RAIN} & \textbf{DD} & \textbf{SLOPE} & \textbf{LST} \\ 
\midrule
PD &	1 &	2 &	4 &	5 &	7 &	9 \\
LULC &	0.5 &	1 &	3 &	4 &	5 &	7 \\
RAIN &	0.25 &	0.33 &	1 &	2 &	4 &	5 \\
DD &	0.2 &	0.25 &	0.5 &	1 &	3 &	5 \\
SLOPE &	0.14 &	0.2 &	0.25 &	0.33 &	1 &	4 \\
LST &	0.11 &	0.14 &	0.2 &	0.2 &	0.25 &	1 \\
\midrule
Total &	2.2 &	3.92 &	8.95 &	12.53 &	20.25 &	31 \\
\bottomrule
\end{tabular} 
\\
\vspace{0.5em}

b.) \textbf{Normalized AHP pairwise comparison matrix}
\begin{tabular}{@{}cccccccc@{}}
\toprule
  & \textbf{PD} & \textbf{LULC} & \textbf{RAIN} & \textbf{DD} & \textbf{SLOPE} & \textbf{LST} & \textbf{Factor Weight} \\ 
\midrule
PD &	0.45 &	0.51 &	0.45 &	0.4 &	0.35 &	0.29 &	0.41 \\
LULC &	0.23 &	0.26 &	0.34 &	0.32 &	0.25 &	0.23 &	0.27 \\
RAIN &	0.11 &	0.08 &	0.11 &	0.16 &	0.2 &	0.16 &	0.14 \\
DD &	0.09 &	0.06 &	0.06 &	0.08 &	0.15 &	0.16 &	0.1 \\
SLOPE &	0.06 &	0.05 &	0.03 &	0.03 &	0.05 &	0.13 &	0.06 \\
LST &	0.05 &	0.04 &	0.02 &	0.02 &	0.01 &	0.03 &	0.03 \\
\bottomrule
\end{tabular}
\vspace{0.5em}
\begin{minipage}{0.75\textwidth}
    \textsuperscript{a}\footnotesize Pairwise comparison matrix for AHP factors, their normalized weight matrix, and overall mean normalized weights which were used to determine a vulnerability score per pixel. PD, population density; LULC, land use and land change; RAIN, rainfall; DD, drainage density; LST, land surface temperature.
    \end{minipage}
\end{adjustwidth}
\end{table}

\paragraph*{S1 Text}
\label{S1_Text}
{\bf AHP Ranking method and pairwise comparisons}

We used the factors Population Density (PD), Land Use Land Cover (LULC), Rainfall (RAIN), drainage density (DD), Slope, and Land Surface Temperature (LST) to define the hierarchy order among parameters on the basis of their importance or relevance in terms of assessing pollution vulnerability. We classified the values for each parameter for each pixel within our area of interest between 1-5 to arrive at the “Factor Class” for that parameter, and we multiplied each Factor Class with the weight indicating the importance of that parameter. We computed the sum of all weighted parameters to produce the overall vulnerability of each pixel. We compared this with the classification parameter ranges as defined using AHP. 

The importance of factors influencing naala vulnerability were placed in the following order: Population Density (PD) > LULC > Rainfall (RAIN) > Drainage Density (DD) > Slope > LST, and the factor importance rankings were inputs into an AHP pairwise comparison matrix (\nameref{S3_Table}).

\nameref{S3_Table} contains the pairwise comparison matrix and the Normalized AHP pairwise comparison matrix. The Principal Eigenvalue is the mean of the eigenvectors, and n is the number of factors - six.

To begin the AHP process, we constructed a pairwise comparison matrix (A), where each element \(a_{ij}\) represented the importance of criterion \(i\) relative to criterion \(j\).


Once we created our pairwise comparison matrix, we normalized the matrix by dividing each element of a column by the sum of the elements of that column. This resulted in the normalized matrix \( \tilde{A} \). The normalized matrix for our example was computed by:

\begin{equation}\label{eq4}
\tilde{a}_{ij} = \frac{a_{ij}}{\sum_{i=1}^{n}a_{ij}}
\end{equation}

The priority vector \(W\) (the eigenvector) was then derived by averaging the rows of the normalized matrix:

\begin{equation}\label{eq5}
w_i = \frac{1}{n} \sum_{j=1}^{n} \tilde{a}_{ij}
\end{equation}

This priority vector defines the AHP weights for each factor, and thus the impact of each factor on the vulnerability of Ganga.

In some situations, AHP importance rankings may produce inconsistent outcomes due to matrix properties. The Consistency Index (\(CI\)) and Consistency Ratio (\(CR\)) were developed to test for inconsistency with a Consistency Ratio < 0.1 indicating the rankings were consistent~\citep{20}:

\begin{equation}\label{eq6}
CI = \frac{\lambda_{max} - n}{n - 1}
\end{equation}

where \( \lambda_{max} \) was the largest eigenvalue of the matrix \(A\). The \(CR\) was the ratio of the \(CI\) to the Random Consistency Index (\(RI\)) for the corresponding matrix size (n):

\begin{equation}\label{eq7}
CR = CI/RI
\end{equation}

In our case, with a consistency ratio (\(CR\)) of 0.06, the matrix was considered consistent (as \(CR\) < 0.1). This consistency indicated that our pairwise comparison matrix was reliable for determining the weights.

\paragraph*{S2 Text}
\label{S2_Text}
{\bf Natural breaks and AHP factor classes}
Data for each AHP factor were divided into five vulnerability classes using the Jenks natural breaks method~\citep{7}. The Jenks method is widely used within GIS packages to generate variance-minimization classification. Breaks were typically uneven, and were selected to separate values where large changes in value occur. For this, we used the boxplots (\nameref{S1_Fig}) and histograms (\nameref{S2_Fig}) for the layers to identify the natural breaks for each factor.

\paragraph*{S1 Fig.}
\begin{center}
    \includegraphics[width=0.7\textwidth]{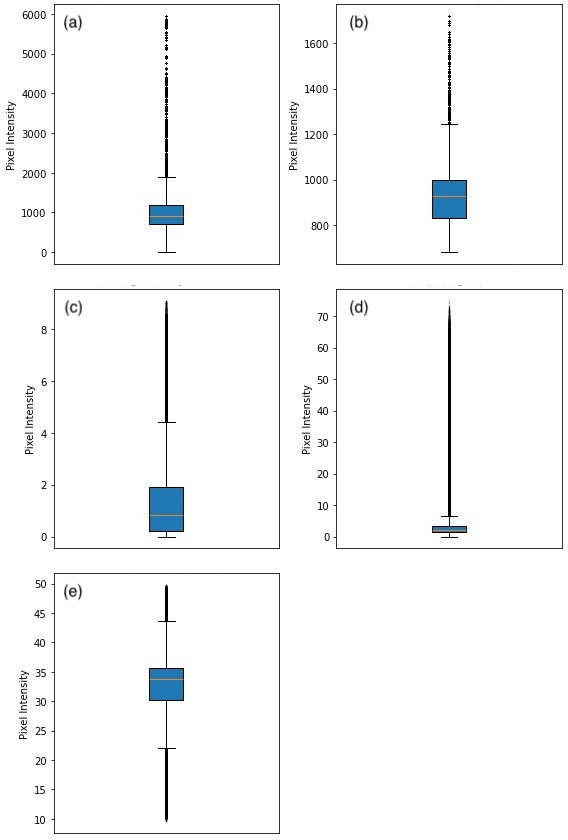}    
\end{center}
\label{S1_Fig}
{\bf AHP Box and Whisker Plot Summary} Panels (a) through (e) are box and whisker plots for the AHP factors Population Density (a), Annual Rainfall (b), Drainage Density (c), Slope (d), and Land Surface Temperature (e) for the length of the Ganga contained in the area of analysis. Data distribution was tested and results are given for each factor label. Whiskers show the range of values, the box "shoulders" are the inter-quartile range, and the median is shown as a band.

\paragraph*{S2 Fig.}
\begin{center}
    \includegraphics[width=0.5\textwidth]{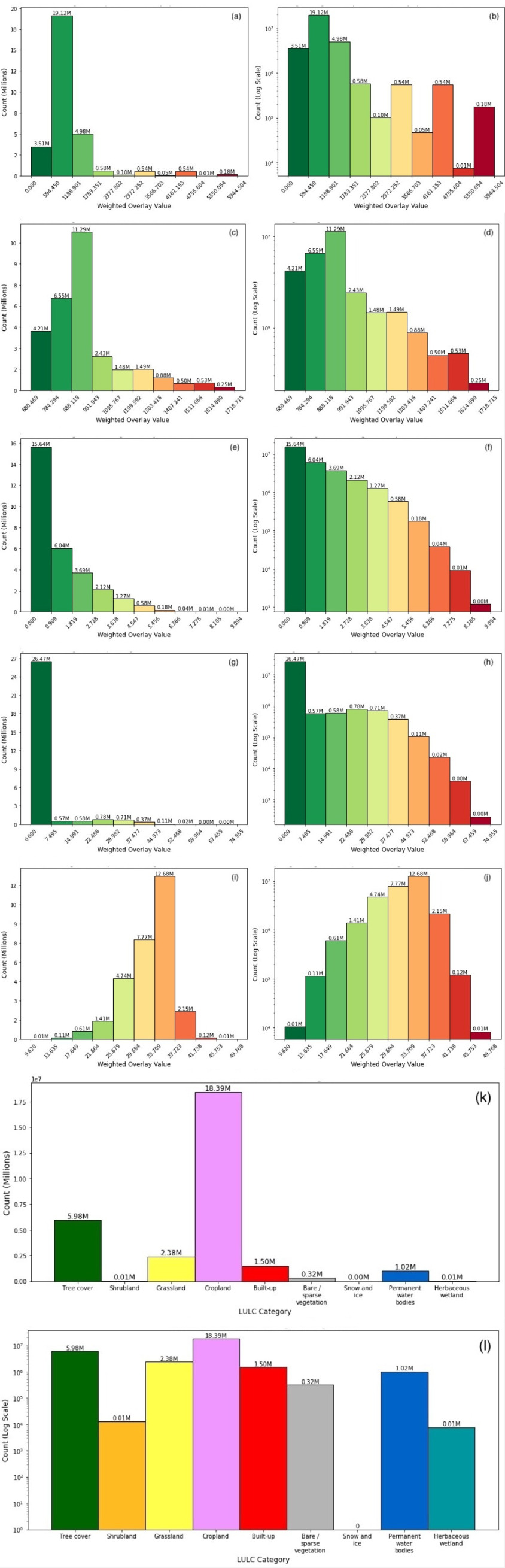}    
\end{center}
\label{S2_Fig}
{\bf AHP Histogram Summary} Panels (a) through (l) are the histograms and log-histograms to assess the distributions of the AHP factors along the length of the Ganga contained in the area of analysis. Panels (a), (c), (e), (g), (i) represent the histograms for Population Density, Annual Rainfall, Drainage Density, Slope, and Land Surface Temperature, respectively, while panels (b), (d), (f), (h), (j) respectively represent the log-histograms for the same.  Panels (a) through (j) are binned using deciles, and coloured green to red for low to high vulnerability. (k) and (l) represent LULC and are color coded to their respective Land Use classes.

\paragraph*{S3 Text}
\label{S3_Text}
{\bf Nested AHP pairwise comparisons and calculation} 

Using a pairwise comparison matrix, we assessed the relationship between individual parameter values. \nameref{S4_Table} shows the pairwise comparison of Population Density and the weight each category has, e.g. PD of >4036 people per km\textsuperscript{2} has a 9 times greater effect on vulnerability than when PD was <865 people per km\textsuperscript{2}. The pairwise matrices for the other factors alongside the explanations of the comparison indices is as follows:

Minimum and maximum values for the five factor classes of population density, slope, rainfall, and drainage density were averaged, normalized, transformed to a 9-point scale to create the factor rankings, and fit to the Saaty AHP scale based on their relationships with pollution production or its transport as well as their breadth across a 9-point scale (~\nameref{S3_Text} and \nameref{S4_Table}). The relationship between population density and pollution production was assumed linear based on municipal solid waste production in India~\citep{S_Goel}, and the factors were ranked as 1, 4, 6, 8, and 9. Rainfall factor rankings spanned 1 through 5 on the Saaty scale because the range between lowest and highest average annual precipitation was a factor of 2 (1, 3, 4, 5, 5). Drainage density can have a nonlinear relationship with pollution transport~\citep{3}, and although the range of drainage density values was more than 20-fold, its factors were ranked as 1, 6, 8, 9, and 9. The relationship between slope and transport is nonlinear for slopes greater than ~26 degrees, and linear for slopes less than 26 degrees~\citep{S_Andrews}. So slope factor rankings were ranked 1, 1.5, 3, 6, and 9. Minimum and maximum temperature data per factor rank were used to calculate normalized temperature dependent microbial growth rates with a Q10 (temperature coefficient) of 1.75. Then the growth rates were transformed to a 9-point scale and scored on the Saaty scale (1, 4, 5, 6, 8). LULC factor ranks were based on ESA WoldCover map~\citep{S_Zanaga} categories (built-up, cropland, bare/sparsely vegetated, tree cover/grassland, water/wetlands) (1,3,4,5,7). Built-up was treated as the factor rank 1 because of the breadth of pollution sources (e.g., municipal solid wastes, pathogens, excess nutrients, and heavy metals) and amount of impervious surfaces that lead to pollution transport. Cropland was ranked 3 due to runoff of fertilizers and agrochemicals. Bare/sparsely vegetated land was ranked 4 due to moderate transport risk. Tree cover and grassland were ranked 5 on the Saaty scale because these categories represent natural ecosystems, and water/wetlands were ranked as 7 on the Saaty scale, because Ganga was the point of pollution deposition.

Running AHP on this pairwise matrix for Population Density gave the subweights for classes 1 - 5. Similarly, once we ran this for the other factors; LULC, Slope, Precipitation, Drainage Density, and Temperature, we got the priority vector of subweights for each factor (\nameref{S4_Table}).

\begin{table}[H]
\centering
\renewcommand{\arraystretch}{1.2} 
\begin{tabular}{|p{4cm}|p{2cm}|c|c|c|c|c|}
\hline
\textbf{Factor} & \textbf{AHP Weight} & \multicolumn{5}{c|}{\textbf{Weight of Vulnerability Classes}} \\ 
\hline
 &  & \textbf{5} & \textbf{4} & \textbf{3} & \textbf{2} & \textbf{1} \\ 
\hline
Population Density    & 0.408 & 0.572 & 0.207 & 0.114 & 0.057 & 0.05  \\ 
\hline
Land Use Land Cover   & 0.268 & 0.505 & 0.186 & 0.144 & 0.110 & 0.055 \\ 
\hline
Slope                 & 0.138 & 0.501 & 0.172 & 0.124 & 0.102 & 0.102 \\ 
\hline
Precipitation         & 0.100 & 0.642 & 0.157 & 0.071 & 0.065 & 0.065 \\ 
\hline
Drainage Density      & 0.058 & 0.416 & 0.278 & 0.202 & 0.070 & 0.034 \\ 
\hline
Temperature           & 0.028 & 0.557 & 0.165 & 0.129 & 0.099 & 0.050 \\ 
\hline
\end{tabular}
\label{table1}
\end{table}

To create the vulnerability map on the Ganga buffer region using Nested AHP, we calculated the predicted vulnerability for each pixel, where:

\begin{itemize}
\item \(W\) is the 1x6 row vector of AHP weights:
\begin{equation}\label{eq8}
W = \begin{bmatrix} w_1 & w_2 & w_3 & w_4 & w_5 & w_6 \end{bmatrix}
\end{equation}
					
\item \(X(p)\) is the 1x6 row vector of parameter values at pixel p for the six factors:
\begin{equation}\label{eq9}
X(p) =\begin{bmatrix} x_1(p) & x_2(p) & x_3(p) & x_4(p) & x_5(p) & x_6(p)
\end{bmatrix}
\end{equation}

\item \(S\) is the 5x6 matrix of subweights, where each column corresponds to a factor and each row corresponds to the subweights for factor classes 1, 2, 3, 4, 5. 

\begin{equation}\label{eq10}
S = \begin{bmatrix} s_1(1) & s_2(1) & s_3(1) & s_4(1) & s_5(1) & s_6(1) \\ s_1(2) & s_2(2) & s_3(2) & s_4(2) & s_5(2) & s_6(2) \\ s_1(3) & s_2(3) & s_3(3) & s_4(3) & s_5(3) & s_6(3) \\ s_1(4) & s_2(4) & s_3(4) & s_4(4) & s_5(4) & s_6(4) \\ s_1(5) & s_2(5) & s_3(5) & s_4(5) & s_5(5) & s_6(5) \end{bmatrix}
\end{equation}
\end{itemize}

Each entry \(s_i(v)\) corresponds to the subweight of factor \(i\) when the value is \(v\).
For each pixel \(p\), the parameter vector \(X(p)\) is used to select corresponding subweights from \(S\) based on its value. This can be done through matrix multiplication by creating a matrix \(P\), which will be a 5x6 binary matrix (one-hot encoded) representing the values of each parameter.
For instance, if \(x_1(p) = 3, x_2(p) = 1\) the matrix \(P\) will have a 1 in the corresponding positions for the values \(x_1(p), x_2(p)\), \ldots at pixel \(p\):

\begin{equation}\label{eq11}
P = \begin{bmatrix} 0 & 1 & 0 & 0 & 0 & 0 \\ 0 & 0 & 0 & 0 & 1 & 0 \\ 1 & 0 & 0 & 0 & 0 & 0 \\ 0 & 0 & 1 & 0 & 0 & 0 \\ 0 & 0 & 0 & 1 & 0 & 1 \end{bmatrix}		\end{equation}

Now, we performed the matrix multiplication between \(S\) (the subweights matrix) and \(P\) (the binary matrix).

\begin{equation}\label{eq12}
S \cdot P = \begin{bmatrix} s_1(x_1(p)) & s_2(x_2(p)) & ... & s_6(x_6(p)) \end{bmatrix}
\end{equation}

This operation selects the subweights corresponding to the actual values of the parameters at pixel p. Finally, we
performed the dot product of the AHP weight vector \(W\) with the transpose of this resulting subweight vector \(S.P\) to obtain the predicted vulnerability for each pixel using Nested AHP \(V(p)\):

\begin{equation}\label{eq13}
V(p) = W \cdot (S \cdot P)^T
\end{equation}

\paragraph*{S4 Table.}
\label{S4_Table}
{\bf Nested AHP: Factor Vulnerability Rankings for all Factors}
\begin{adjustwidth}{-2.25in}{0in}
\centering
\renewcommand{\arraystretch}{1.2} 
\begin{tabular}{@{}p{4cm}p{1cm}p{1.25cm}p{2.5cm}p{2cm}p{2cm}@{}}
    \toprule
    \textbf{LULC} &	\textbf{Built-up} &	\textbf{Cropland} &	\textbf{Barren/ Sparse Vegetation} &	\textbf{Trees/ Grass/ Shrubs} &	\textbf{Water/ Wetland} \\
    \midrule
    Built-up &	1 &	3 &	4 &	5 &	7 \\
    Cropland &	0.33 &	1 &	1 &	2 &	4 \\
    Barren/Sparse Vegetation  &	0.25 &	1 &	1 &	1 &	3 \\
    Trees/ Grass/ Shrubs  &	0.2 &	0.5 &	1 &	1 &	2 \\
    Water / Wetland &	0.14 &	0.25 &	0.33 &	0.5 &	1 \\
    \midrule
    Total &	1.93 &	5.75 &	7.33 &	10 &	17 \\
    \bottomrule
\end{tabular}
\begin{tabular}{@{}p{2.5cm}p{1cm}p{1cm}p{1cm}p{1cm}p{1cm}@{}}
    \toprule
    \textbf{Rainfall (mm/y)} &	\textbf{>1410} &	\textbf{1189-1410} &	\textbf{1019-1189} &	\textbf{868-1019} &	\textbf{<868} \\
    \midrule
    >1410 &	1 &	3 &	4 &	5 &	5 \\
    1189-1410 &	0.33 &	1 &	1 &	2 &	2 \\
    1019-1189 &	0.25 &	1 &	1 &	1 &	1 \\
    868-1019 &	0.2 &	0.5 &	1 &	1 &	1 \\
    <868 &	0.2 &	0.5 &	1 &	1 &	1 \\
    \midrule
    Total &	1.98 &	6 &	8 &	10 & 10 \\
    \bottomrule
\end{tabular}
\begin{tabular}{@{}p{2.5cm}p{1cm}p{1cm}p{1cm}p{1cm}p{1cm}@{}}
    \toprule
    \textbf{Drainage density (km/km\textsuperscript{2})} &	\textbf{>4} &	\textbf{2.6-4} &	\textbf{1.5-2.6} &	\textbf{0.6-1.5} &	\textbf{<0.6} \\
    \midrule
    >4 &	1 &	6 &	8 &	9 &	9 \\
    2.6-4 &	0.17 &	1 &	2 &	3 &	3 \\
    1.5-2.6 &	0.13 &	0.5 &	1 &	1 &	1 \\
    0.6-1.5 &	0.11 &	0.33 &	1 &	1 &	1 \\
    <0.6 &	0.11 &	0.33 &	1 &	1 &	1 \\
    \midrule
    Total &	1.51 &	8.17 &	13 &	15 &	15 \\
    \bottomrule
\end{tabular}
\begin{tabular}{@{}p{2.5cm}p{1cm}p{1cm}p{1cm}p{1cm}p{1cm}@{}}
    \toprule
    \textbf{Slope (Angle \%)} &	\textbf{>35} &	\textbf{24-35} &	\textbf{45650} &	\textbf{45394} &	\textbf{<4} \\
    \midrule
    >35 &	1 &	1.5 &	3 &	6 &	9 \\
    24-35 &	0.67 &	1 &	2 &	5 &	8 \\
    45650 &	0.33 &	0.67 &	1 &	4 &	7 \\
    45394 &	0.17 &	0.22 &	0.25 &	1 &	3 \\
    <4 &	0.11 &	0.13 &	0.14 &	0.33 &	1 \\
    \midrule
    Total &	2.28 &	3.52 &	5.89 &	16 &	28 \\
    \bottomrule
\end{tabular}
\begin{tabular}{@{}p{2.5cm}p{1cm}p{1cm}p{1cm}p{1cm}p{1cm}@{}}
    \toprule
    \textbf{Temperature (℃)} &	\textbf{>37} &	\textbf{34-37} &	\textbf{30-34} &	\textbf{25-30} &	\textbf{<25} \\
    \midrule
    >37 &	1 &	4 &	5 &	6 &	8 \\
    34-37 &	0.25 &	1 &	1 &	2 &	4 \\
    30-34 &	0.2 &	1 &	1 &	1 &	3 \\
    25-30 &	0.17 &	0.5 &	1 &	1 &	2 \\
    <25 &	0.13 &	0.25 &	0.33 &	0.5 &	1 \\
    \midrule
    Total &	1.74 &	6.75 &	8.33 &	11 &	18 \\
    \bottomrule
\end{tabular}
\end{adjustwidth}

\paragraph*{S4 Text}
\label{S4_Text}
{\bf Fuzzy AHP calculation}

In Fuzzy AHP, preferences are represented as Triangular Fuzzy Numbers (TFNs) rather than single discrete values. A triplet \((l,m,u)\) is known as a Triangular Fuzzy Number \(\tilde{A}\), where l represents the lower bound, m is the most likely value, and u is the upper bound. These numbers form a triangular shape on a graph, and they allow factors to have a range of possible values for each pairwise comparison, rather than a single exact number. For example, instead of stating that one criterion is exactly twice as important as another, a factor can be expressed such that it is “between 1.5 and 2.5 times more important, with 2 being the most likely value.” This flexibility helps to capture the subjective uncertainty inherent in human judgments.

Once we used these nonlinear weightages in our AHP calculations, we updated the map showing the different classifications of pixels based on vulnerability.

The Fuzzy Pairwise Comparison Matrix was constructed using these Triangular Fuzzy Numbers (TFN). Each element \(a~ij\tilde{a}_{ij}a~ij\) in the matrix was a TFN representing the relative importance of criterion i over criterion j. If a criterion was compared with itself, its TFN was (1,1,1), representing absolute equality. For off-diagonal elements, if \(\tilde{a}_{ij} = (l, m, u)\), then \(\tilde{a}_{ji} = \left(\frac{1}{u}, \frac{1}{m}, \frac{1}{l}\right)\), maintaining the reciprocal nature of the matrix. These fuzzy comparisons were then aggregated to derive the fuzzy weight vector for each criterion.

In Fuzzy AHP, the decision maker generally recreates the pairwise matrix by choosing the \(l, m, u\) for each element, but to understand the outlying cases for Fuzzy AHP, and to incorporate the breadth of decision maker’s range, we ran Fuzzy AHP on 100,000 Monte Carlo simulations at 95\% fuzziness. For each run, we made a unique Fuzzy Pairwise Comparison Matrix, where if the comparison index between two factors was 2, the triangular fuzzy range would be \((0.05*2, 2, 2*1.95) = (0.1, 2, 3.9)\), and we chose a random value between 0.1 and 2, and 2 and 3.9 to design the final TFN \([random.uniform(0.1,2), 2, random.uniform(2, 3.9)]\). Running 100,000 simulations took into account most of the likely Fuzzy Pairwise Comparison Matrices that might be generated. 

We ran Fuzzy AHP on each Fuzzy Pairwise Comparison Matrix, where we calculated the geometric mean across each row of the matrix, and normalized the results. The final step was to defuzzify these fuzzy numbers by taking the mean, which returned the priority vector.
We then determined the number of cases out of 100,000 where there were rank reversals. We observed the cases where the order of parameters in terms of vulnerability changed, as compared to rank-orders in case of classic AHP.

\paragraph*{S5 Text}
\label{S5_Text}
{\bf ANP supermatrix and inner dependence (W22 and W33) pairwise comparisons}
Pairwise comparisons were made to describe the relative importance of the ANP criteria (AHP factors) on influencing each other (\nameref{S3_Fig}). Criteria that did not contribute to an inner dependence matrix were scored with a zero. When identifying the relative importance of Land Use Land Cover (LULC) compared to other criteria on influencing Population Density (PD), it was scored with a 2 because of the importance of built-up and agriculture categories in LULC and their direct relationships with population densities. Rainfall and slope were scored with a 5 because historically there were natural limits to both that influenced urbanization and agriculture although this is changing~\citep{S_Zhou, S_Chen, S_Shi, S_Wang, S_Ullah}. Drainage Density (DD) was scored an 8 because the natural drainage density is modified by people to meet agricultural and societal requirements suggesting it has a minor influence on PD. When identifying the relative importance of PD compared to other criteria on influencing LULC, it was scored with a 2 as a reciprocal of the LULC-PD inner dependence. Rainfall and temperature are major drivers of land cover type, and as such each was scored with a 4. Slope and DD shape landscapes within land cover types, and each was scored with a 7. When identifying the relative importance of LULC compared to other criteria on influencing rainfall, it was scored with a 7 and PD was scored with an 8 because of the potential for built-up land to influence local precipitation~\citep{S_Liu}. Slope may influence the speed of runoff during rainfall and was scored with a 5. When identifying the relative importance of criteria on influencing DD, PD was scored a 2, LULC was scored a 4, and slope was scored a 7, because DD is the product of both anthropogenic and natural influences. When identifying the relative importance of criteria on influencing temperature, PD was scored a 5, and LULC was scored an 8 to account for potential urban heat island effects.

\paragraph*{S3 Fig.}
\begin{center}
    \includegraphics[width=0.7\textwidth]{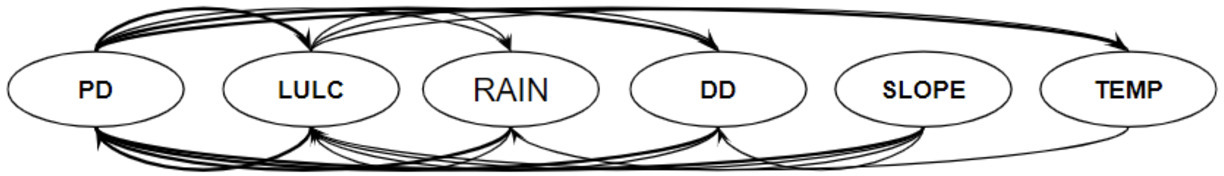}    
\end{center}
\label{S3_Fig}
{\bf ANP Parameter Inner Dependencies} Inner dependencies between ANP criteria (AHP factors) are shown as ovals. Arrows about the ovals indicate influence by a criterion on other criteria to its right and arrows below the ovals indicate influence by a criterion on other criteria to its left.

\paragraph*{S6 Text}
\label{S6_Text}
{\bf 1-N AHP calculation}

We first took our original output of AHP, i.e., the weights of the known factors. Here, we assumed that our known factors would at least represent 67\% of the importance while classifying the vulnerability variable. This is a fair assumption since through our preliminary analysis we can conclude that our factors were robust enough to represent at least 67\% of the variability of the vulnerability variable.

The unknown factors could be one of two types: acute or chronic. Some factors (acute) may be less likely to occur but more likely to have a significant impact on vulnerability, while other factors (chronic) may be more likely to occur but less likely to have a less significant impact. In the case of our problem statement, acute factors might include mining sites and hide tanning operations while chronic factors might include algal blooms from fertilizer runoff and pollution during funeral processions.

To assess the potential impact of these acute and chronic factors, we first created two new layers on the river Ganga buffer region, with uniformly distributed random values from 1-5 similar to the other factors, and pixels with value zero, to capture the probabilistic non-occurrence of acute or chronic factors. For chronic factors, 75\% of the pixels were 0, while for acute factors 97.5\% of the pixels were 0. This means that each variable had a 25\% and a 2.5\% probability of occurrence respectively, where pixels were assigned the random values 1-5. 

To evaluate the unknown variability captured between 1-N (where N lies between 0.67 and 1), we split it between acute and chronic. We assumed that since acute factors have a significant impact on vulnerability, 90\% of the 1-N factor would be attributed to the impact of acute factors to vulnerability, while the leftover 10\% would be attributed to the impact of chronic factors to vulnerability. For instance, in the worst case scenario if 1-N is 0.33 (i.e., the weightage of the unknown factors towards vulnerability is 33\%), the 0.33 will be split into 0.297 for acute and 0.033 for chronic. The original six factors, with their weights 0.408, 0.268, 0.138, 0.100, 0.058, 0.028 will be scaled down based on N. If N is 0.67, we multiply the old weights by 0.67 so that they sum up to n, and we can accommodate the new acute and chronic weights. Here, 0.408*0.67 gives 0.27336, which would be lower than 0.297 for acute factors, which means the acute factors would have a higher impact on vulnerability but with a lower probability.

We ran AHP using the original six layers and two additional layers, with new weights to accommodate for acute and chronic factors, and compared the results with the AHP results to check the robustness of our model. We ran this approach once with 1-N as 0.33 to represent a worst case scenario, and once with 1-N as 0.165 to reflect the average or expected case. 

To ensure consistent results while dealing with abstract foundations for parameter understanding, we performed our analysis on a set of assumptions. We assumed that our classic AHP model at least represented 67\% of the importance in terms of weights and that the value of 1-N could be anything between 0 and 0.33 with a uniform distribution. Also, to deal with the uncertainty, we assumed that the unknown factors can be of two types - acute and chronic, with a 2.5\% and a 25\% probability of occurrence respectively; and while there could be numerous unknown factors, they can be clumped into two factors - acute and chronic. Lastly, we assumed that the non-zero pixel values were randomly and uniformly distributed with their appropriate probabilities.

\paragraph*{S4 Fig.}
\begin{center}
    \includegraphics[width=0.7\textwidth]{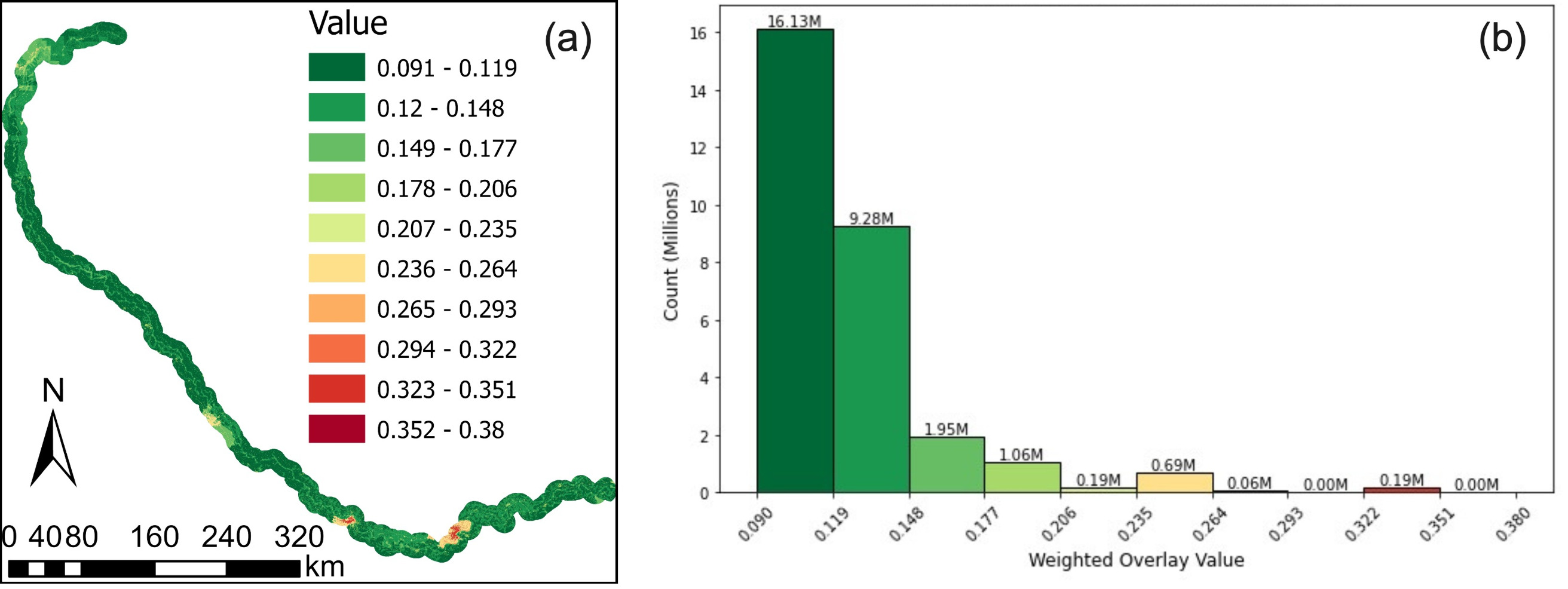}    
\end{center}
\label{S4_Fig}
{\bf ANP with original AHP submatrix} These figures show the ANP’s vulnerability score with the original AHP submatrix; the lower values pertained to the least vulnerable areas while the ones with higher scores were the most vulnerable. Map of exact values per pixel location (a) and histogram of binned ranges of values (b).

\paragraph*{S5 Fig.}
\begin{center}
    \includegraphics[width=0.7\textwidth]{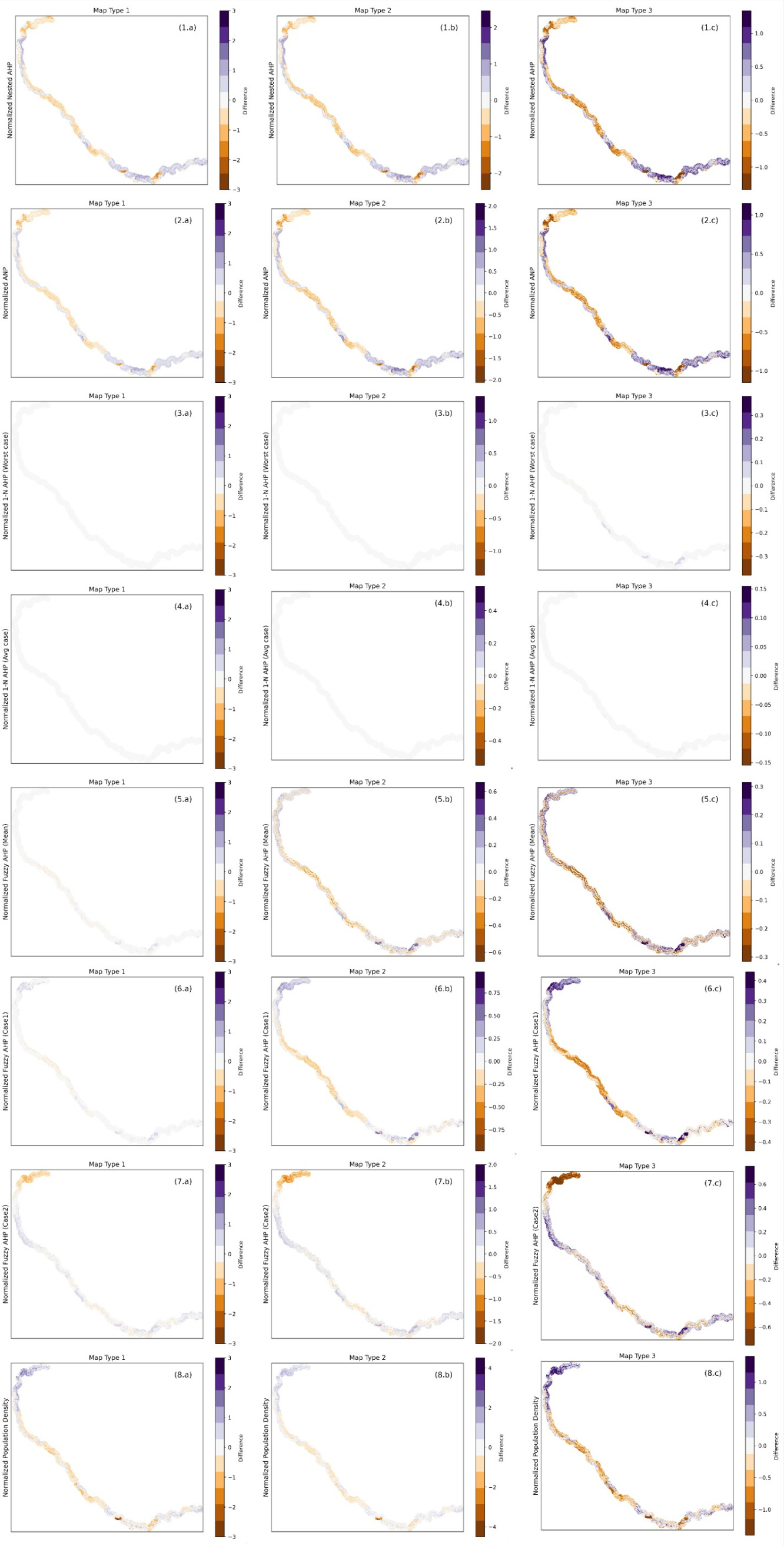}    
\end{center}
\label{S5_Fig}
{\bf Original AHP \& AHP Variants Comparison Chart} These figures show the comparison charts for the different variants of AHP with the original AHP across the Ganga buffer. We used three map types to highlight the distribution of differences between the approaches. Column (a) is Map type 1 (colors spread between overall min max), Column (b) is Map type 2 (colors spread between local min max), and Column (c) is Map type 3 (colors spread between 5-95 percentile). Rows 1-8 are Nested AHP, ANP, 1-N AHP (Worst case), 1-N AHP (Avg case), Fuzzy AHP (Mean), Fuzzy AHP (Case1), Fuzzy AHP (Case2) and Population density, respectively.

\paragraph*{S7 Text}
\label{S7_Text}
{\bf Difference between Fuzzy 1-N AHP, Fuzzy AHP and 1-N AHP}

While Fuzzy 1-N AHP gave us some insights on the combined effects of Fuzzy AHP and 1-N AHP, to test whether it is greater than the sum of its parts, which would warrant an additional comparison on its value addition to AHP, we ran a two-way ANOVA test with the parameters quintile and model type (Fuzzy 1-N AHP, Mean Fuzzy AHP, 1-N AHP Worst case) (\nameref{S7_Table}). While the p-values for quintiles, and the combined effect of quintile and model type on the vulnerability scores were <0.0001, the p-value for models only was 0.9270. When we performed a Tukey post-hoc test for each quintile, we found that (other than in quintile 1) there was no significant difference between Fuzzy 1-N AHP and Fuzzy AHP. Quintiles 1, 4, and 5 of the Fuzzy 1-N AHP and 1-N AHP comparison  were significantly different which reinforced that the edges of the distribution were susceptible to variability caused by uncertainty in model factors as was seen in the 1-N AHP and AHP comparison. Together, these results suggested that the individual analyses of Fuzzy AHP and 1-N AHP can represent the effect of Fuzzy 1-N AHP in comparison with classic AHP. Hence we did not further include Fuzzy 1-N AHP in the comparison analysis or \nameref{S5_Fig}.

\paragraph*{S5 Table.}
\label{S5_Table}
{\bf Two-way ANOVA and Tukey Post-Hoc test for AHP, ANP and Nested AHP}
\begin{adjustwidth}{-2.25in}{0in}
\centering
\begin{tabular}{lcccc}
\toprule
 & \textbf{Sum of Squares} & \textbf{DF} & \textbf{F statistic} & \textbf{p-value} \\
\midrule

Quintile & 18554.85 & 4 & 6539.64 & $<0.0001$ \\ 
Model & 0.51 & 2 & 0.36 & 0.6977702 \\ 
Quintile + Model & 1065.47 & 8 & 187.76 & $<0.0001$ \\ 
Residual & 10629.19 & 14985 & --- & --- \\ 

\bottomrule
\end{tabular}

\vspace{0.5cm}

\begin{tabular}{llccccc}
\toprule
\multicolumn{7}{c}{\textbf{Quintile 1}} \\
\midrule
\textbf{Group 1} & \textbf{Group 2} & \textbf{Mean Diff} & \textbf{p-adj} & \textbf{Lower} & \textbf{Upper} & \textbf{Reject} \\
\midrule

AHP & ANP & 0.458 & $<0.0001$ & 0.3748 & 0.5412 & TRUE \\
AHP & Nested AHP & 0.781 & $<0.0001$ & 0.6978 & 0.8642 & TRUE \\
ANP & Nested AHP & 0.323 & $<0.0001$ & 0.2398 & 0.4062 & TRUE \\

\bottomrule
\end{tabular}

\begin{tabular}{llccccc}
\toprule
\multicolumn{7}{c}{\textbf{Quintile 2}} \\
\midrule
\textbf{Group 1} & \textbf{Group 2} & \textbf{Mean Diff} & \textbf{p-adj} & \textbf{Lower} & \textbf{Upper} & \textbf{Reject} \\
\midrule

AHP & ANP & 0.506 & $<0.0001$ & 0.3949 & 0.6171 & TRUE \\
AHP & Nested AHP & 0.748 & $<0.0001$ & 0.6369 & 0.8591 & TRUE \\
ANP & Nested AHP & 0.242 & $<0.0001$ & 0.1309 & 0.3531 & TRUE \\

\bottomrule
\end{tabular}

\begin{tabular}{llccccc}
\toprule
\multicolumn{7}{c}{\textbf{Quintile 3}} \\
\midrule
\textbf{Group 1} & \textbf{Group 2} & \textbf{Mean Diff} & \textbf{p-adj} & \textbf{Lower} & \textbf{Upper} & \textbf{Reject} \\
\midrule

AHP & ANP & -0.161 & 0.0005 & -0.2616 & -0.0604 & TRUE \\
AHP & Nested AHP & -0.427 & $<0.0001$ & -0.5276 & -0.3264 & TRUE \\
ANP & Nested AHP & -0.266 & $<0.0001$ & -0.3666 & -0.1654 & TRUE \\

\bottomrule
\end{tabular}

\begin{tabular}{llccccc}
\toprule
\multicolumn{7}{c}{\textbf{Quintile 4}} \\
\midrule
\textbf{Group 1} & \textbf{Group 2} & \textbf{Mean Diff} & \textbf{p-adj} & \textbf{Lower} & \textbf{Upper} & \textbf{Reject} \\
\midrule

AHP & ANP & -0.471 & $<0.0001$ & -0.5531 & -0.3889 & TRUE \\
AHP & Nested AHP & -0.668 & $<0.0001$ & -0.7501 & -0.5859 & TRUE \\
ANP & Nested AHP & -0.197 & $<0.0001$ & -0.2791 & -0.1149 & TRUE \\

\bottomrule
\end{tabular}

\begin{tabular}{llccccc}
\toprule
\multicolumn{7}{c}{\textbf{Quintile 5}} \\
\midrule
\textbf{Group 1} & \textbf{Group 2} & \textbf{Mean Diff} & \textbf{p-adj} & \textbf{Lower} & \textbf{Upper} & \textbf{Reject} \\
\midrule

AHP & ANP & -0.385 & $<0.0001$ & -0.4387 & -0.3313 & TRUE \\
AHP & Nested AHP & -0.502 & $<0.0001$ & -0.5557 & -0.4483 & TRUE \\
ANP & Nested AHP & -0.117 & $<0.0001$ & -0.1707 & -0.0633 & TRUE \\

\bottomrule
\end{tabular}
\end{adjustwidth}

\paragraph*{S6 Table.}
\label{S6_Table}
{\bf Two-way ANOVA and Tukey Post-Hoc test for AHP, Mean Fuzzy AHP, Fuzzy AHP Case 1 and Fuzzy AHP Case 2}
\begin{adjustwidth}{-2.25in}{0in}

\centering
\begin{tabular}{lcccc}
\toprule
 & \textbf{Sum of Squares} & \textbf{DF} & \textbf{F statistic} & \textbf{p-value} \\
\midrule

Quintile & 31324.9 & 4 & 18828.71 & $<0.0001$ \\ 
Model & 7.94 & 3 & 6.36 & $<0.0001$ \\ 
Quintile + Model & 534.84 & 12 & 107.16 & $<0.0001$ \\ 
Residual & 8310.07 & 19980 & --- & --- \\ 

\bottomrule
\end{tabular}

\vspace{0.5cm}

\begin{tabular}{llccccc}
\toprule
\multicolumn{7}{c}{\textbf{Quintile 1}} \\
\midrule
\textbf{Group 1} & \textbf{Group 2} & \textbf{Mean Diff} & \textbf{p-adj} & \textbf{Lower} & \textbf{Upper} & \textbf{Reject} \\
\midrule

AHP & Mean Fuzzy AHP & 0.037 & 0.3115 & -0.0182 & 0.0922 & FALSE \\
AHP & Fuzzy AHP Case 1 & 0.28 & $<0.0001$ & 0.2248 & 0.3352 & TRUE \\
AHP & Fuzzy AHP Case 2 & 0.549 & $<0.0001$ & 0.4938 & 0.6042 & TRUE \\
Mean Fuzzy AHP & Fuzzy AHP Case 1 & 0.243 & $<0.0001$ & 0.1878 & 0.2982 & TRUE \\
Mean Fuzzy AHP & Fuzzy AHP Case 2 & 0.512 & $<0.0001$ & 0.4568 & 0.5672 & TRUE \\
Fuzzy AHP Case 1 & Fuzzy AHP Case 2 & 0.269 & $<0.0001$ & 0.2138 & 0.3242 & TRUE \\

\bottomrule
\end{tabular}

\begin{tabular}{llccccc}
\toprule
\multicolumn{7}{c}{\textbf{Quintile 2}} \\
\midrule
\textbf{Group 1} & \textbf{Group 2} & \textbf{Mean Diff} & \textbf{p-adj} & \textbf{Lower} & \textbf{Upper} & \textbf{Reject} \\
\midrule

AHP & Mean Fuzzy AHP & 0.095 & 0.0275 & 0.0073 & 0.1827 & TRUE \\
AHP & Fuzzy AHP Case 1 & 0.04 & 0.6441 & -0.0477 & 0.1277 & FALSE \\
AHP & Fuzzy AHP Case 2 & 0.304 & $<0.0001$ & 0.2163 & 0.3917 & TRUE \\
Mean Fuzzy AHP & Fuzzy AHP Case 1 & -0.055 & 0.3715 & -0.1427 & 0.0327 & FALSE \\
Mean Fuzzy AHP & Fuzzy AHP Case 2 & 0.209 & $<0.0001$ & 0.1213 & 0.2967 & TRUE \\
Fuzzy AHP Case 1 & Fuzzy AHP Case 2 & 0.264 & $<0.0001$ & 0.1763 & 0.3517 & TRUE \\

\bottomrule
\end{tabular}

\begin{tabular}{llccccc}
\toprule
\multicolumn{7}{c}{\textbf{Quintile 3}} \\
\midrule
\textbf{Group 1} & \textbf{Group 2} & \textbf{Mean Diff} & \textbf{p-adj} & \textbf{Lower} & \textbf{Upper} & \textbf{Reject} \\
\midrule

AHP & Mean Fuzzy AHP & 0.053 & 0.3898 & -0.0332 & 0.1392 & FALSE \\
AHP & Fuzzy AHP Case 1 & 0.045 & 0.5361 & -0.0412 & 0.1312 & FALSE \\
AHP & Fuzzy AHP Case 2 & -0.117 & 0.0028 & -0.2032 & -0.0308 & TRUE \\
Mean Fuzzy AHP & Fuzzy AHP Case 1 & -0.008 & 0.9952 & -0.0942 & 0.0782 & FALSE \\
Mean Fuzzy AHP & Fuzzy AHP Case 2 & -0.17 & $<0.0001$ & -0.2562 & -0.0838 & TRUE \\
Fuzzy AHP Case 1 & Fuzzy AHP Case 2 & -0.162 & $<0.0001$ & -0.2482 & -0.0758 & TRUE \\

\bottomrule
\end{tabular}

\begin{tabular}{llccccc}
\toprule
\multicolumn{7}{c}{\textbf{Quintile 4}} \\
\midrule
\textbf{Group 1} & \textbf{Group 2} & \textbf{Mean Diff} & \textbf{p-adj} & \textbf{Lower} & \textbf{Upper} & \textbf{Reject} \\
\midrule

AHP & Mean Fuzzy AHP & -0.129 & 0.0002 & -0.2097 & -0.0483 & TRUE \\
AHP & Fuzzy AHP Case 1 & -0.209 & $<0.0001$ & -0.2897 & -0.1283 & TRUE \\
AHP & Fuzzy AHP Case 2 & -0.541 & $<0.0001$ & -0.6217 & -0.4603 & TRUE \\
Mean Fuzzy AHP & Fuzzy AHP Case 1 & -0.08 & 0.0529 & -0.1607 & 0.0007 & FALSE \\
Mean Fuzzy AHP & Fuzzy AHP Case 2 & -0.412 & $<0.0001$ & -0.4927 & -0.3313 & TRUE \\
Fuzzy AHP Case 1 & Fuzzy AHP Case 2 & -0.332 & $<0.0001$ & -0.4127 & -0.2513 & TRUE \\

\bottomrule
\end{tabular}

\begin{tabular}{llccccc}
\toprule
\multicolumn{7}{c}{\textbf{Quintile 5}} \\
\midrule
\textbf{Group 1} & \textbf{Group 2} & \textbf{Mean Diff} & \textbf{p-adj} & \textbf{Lower} & \textbf{Upper} & \textbf{Reject} \\
\midrule

AHP & Mean Fuzzy AHP & -0.104 & $<0.0001$ & -0.157 & -0.051 & TRUE \\
AHP & Fuzzy AHP Case 1 & -0.227 & $<0.0001$ & -0.28 & -0.174 & TRUE \\
AHP & Fuzzy AHP Case 2 & -0.457 & $<0.0001$ & -0.51 & -0.404 & TRUE \\
Mean Fuzzy AHP & Fuzzy AHP Case 1 & -0.123 & $<0.0001$ & -0.176 & -0.07 & TRUE \\
Mean Fuzzy AHP & Fuzzy AHP Case 2 & -0.353 & $<0.0001$ & -0.406 & -0.3 & TRUE \\
Fuzzy AHP Case 1 & Fuzzy AHP Case 2 & -0.23 & $<0.0001$ & -0.283 & -0.177 & TRUE \\

\bottomrule
\end{tabular}
\end{adjustwidth}

\paragraph*{S7 Table.}
\label{S7_Table}
{\bf Two-way ANOVA and Tukey Post-Hoc test for Fuzzy 1-N AHP, Fuzzy AHP and 1-N AHP}
\begin{adjustwidth}{-2.25in}{0in}
\centering
\begin{tabular}{lcccc}
\toprule
 & \textbf{Sum of Squares} & \textbf{DF} & \textbf{F statistic} & \textbf{p-value} \\
\midrule

Quintile & 24588.76 & 4 & 17336.98 & $<0.0001$ \\ 
Model & 0.05 & 2 & 0.08 & 0.9270276 \\ 
Quintile + Model & 109.21 & 8 & 38.5 & $<0.0001$ \\ 
Residual & 5313.25 & 14985 & --- & --- \\ 

\bottomrule
\end{tabular}

\vspace{0.5cm}

\begin{tabular}{llccccc}
\toprule
\multicolumn{7}{c}{\textbf{Quintile 1}} \\
\midrule
\textbf{Group 1} & \textbf{Group 2} & \textbf{Mean Diff} & \textbf{p-adj} & \textbf{Lower} & \textbf{Upper} & \textbf{Reject} \\
\midrule

Fuzzy 1-N AHP & Fuzzy AHP & -0.096 & $<0.0001$ & -0.1433 & -0.0487 & TRUE \\
Fuzzy 1-N AHP & 1-N AHP & 0.27 & $<0.0001$ & 0.2227 & 0.3173 & TRUE \\
Fuzzy AHP & 1-N AHP & 0.366 & $<0.0001$ & 0.3187 & 0.4133 & TRUE \\

\bottomrule
\end{tabular}

\begin{tabular}{llccccc}
\toprule
\multicolumn{7}{c}{\textbf{Quintile 2}} \\
\midrule
\textbf{Group 1} & \textbf{Group 2} & \textbf{Mean Diff} & \textbf{p-adj} & \textbf{Lower} & \textbf{Upper} & \textbf{Reject} \\
\midrule

Fuzzy 1-N AHP & Fuzzy AHP & 0.053 & 0.1918 & -0.0186 & 0.1246 & FALSE \\
Fuzzy 1-N AHP & 1-N AHP & 0.055 & 0.169 & -0.0166 & 0.1266 & FALSE \\
Fuzzy AHP & 1-N AHP & 0.002 & 0.9976 & -0.0696 & 0.0736 & FALSE \\

\bottomrule
\end{tabular}

\begin{tabular}{llccccc}
\toprule
\multicolumn{7}{c}{\textbf{Quintile 3}} \\
\midrule
\textbf{Group 1} & \textbf{Group 2} & \textbf{Mean Diff} & \textbf{p-adj} & \textbf{Lower} & \textbf{Upper} & \textbf{Reject} \\
\midrule

Fuzzy 1-N AHP & Fuzzy AHP & 0.072 & 0.0682 & -0.0041 & 0.1481 & FALSE \\
Fuzzy 1-N AHP & 1-N AHP & -0.007 & 0.9747 & -0.0831 & 0.0691 & FALSE \\
Fuzzy AHP & 1-N AHP & -0.079 & 0.0397 & -0.1551 & -0.0029 & TRUE \\

\bottomrule
\end{tabular}

\begin{tabular}{llccccc}
\toprule
\multicolumn{7}{c}{\textbf{Quintile 4}} \\
\midrule
\textbf{Group 1} & \textbf{Group 2} & \textbf{Mean Diff} & \textbf{p-adj} & \textbf{Lower} & \textbf{Upper} & \textbf{Reject} \\
\midrule

Fuzzy 1-N AHP & Fuzzy AHP & -0.001 & 0.9993 & -0.0681 & 0.0661 & FALSE \\
Fuzzy 1-N AHP & 1-N AHP & -0.099 & 0.0016 & -0.1661 & -0.0319 & TRUE \\
Fuzzy AHP & 1-N AHP & -0.098 & 0.0018 & -0.1651 & -0.0309 & TRUE \\

\bottomrule
\end{tabular}

\begin{tabular}{llccccc}
\toprule
\multicolumn{7}{c}{\textbf{Quintile 5}} \\
\midrule
\textbf{Group 1} & \textbf{Group 2} & \textbf{Mean Diff} & \textbf{p-adj} & \textbf{Lower} & \textbf{Upper} & \textbf{Reject} \\
\midrule

Fuzzy 1-N AHP & Fuzzy AHP & -0.009 & 0.8753 & -0.0519 & 0.0339 & FALSE \\
Fuzzy 1-N AHP & 1-N AHP & -0.198 & $<0.0001$ & -0.2409 & -0.1551 & TRUE \\
Fuzzy AHP & 1-N AHP & -0.189 & $<0.0001$ & -0.2319 & -0.1461 & TRUE \\

\bottomrule
\end{tabular}
\end{adjustwidth}

\end{document}